\documentclass[useAMS,usenatbib]{mn2e}
\pdfoutput=1
\usepackage{graphicx}    % Include figure files
\usepackage{dcolumn}     % Align table columns on decimal point
\usepackage{bm}          % bold math
\usepackage{amssymb,amsmath}  
\usepackage{color}
\usepackage{overpic}
\usepackage[colorinlistoftodos]{todonotes}
\usepackage[draft]{hyperref}
\usepackage{lineno}
\title[Measuring galaxy bias with CiC]
      {Measuring Linear and Non-linear Galaxy Bias Using Counts-in-Cells in the Dark Energy Survey Science Verification Data}
\author[A.~I.~Salvador, F.~J.~S\'anchez et al.]{
\parbox{\textwidth}{
\Large
A.~I.~Salvador$^{1}$\thanks{Email:ana.salvador@uam.es},
F.~J.~S\'anchez$^{2,3}$\thanks{Email:francs1@uci.edu},
A.~Pagul$^{4,5}$,
J.~Garc\'ia-Bellido$^{1}$,
E.~Sanchez$^{3}$,
A.~Pujol$^{6,7}$,
J.~Frieman$^{8,5}$,
E.~Gaztanaga$^{9,10}$,
A.~J.~Ross$^{11}$,
I.~Sevilla-Noarbe$^{3}$,
T.~M.~C.~Abbott$^{12}$,
S.~Allam$^{8}$,
J.~Annis$^{8}$,
S.~Avila$^{13}$,
E.~Bertin$^{14,15}$,
D.~Brooks$^{16}$,
D.~L.~Burke$^{17,18}$,
A.~Carnero~Rosell$^{19,20}$,
M.~Carrasco~Kind$^{21,22}$,
J.~Carretero$^{23}$,
F.~J.~Castander$^{9,10}$,
C.~E.~Cunha$^{17}$,
J.~De~Vicente$^{3}$,
H.~T.~Diehl$^{8}$,
P.~Doel$^{16}$,
A.~E.~Evrard$^{24,25}$,
P.~Fosalba$^{9,10}$,
D.~Gruen$^{17,18}$,
R.~A.~Gruendl$^{21,22}$,
J.~Gschwend$^{19,20}$,
G.~Gutierrez$^{8}$,
W.~G.~Hartley$^{16,26}$,
D.~L.~Hollowood$^{27}$,
D.~J.~James$^{28}$,
K.~Kuehn$^{29}$,
N.~Kuropatkin$^{8}$,
O.~Lahav$^{16}$,
M.~Lima$^{30,19}$,
M.~March$^{31}$,
J.~L.~Marshall$^{32}$,
F.~Menanteau$^{21,22}$,
R.~Miquel$^{33,23}$,
A.~K.~Romer$^{34}$,
A.~Roodman$^{17,18}$,
V.~Scarpine$^{8}$,
R.~Schindler$^{18}$,
M.~Smith$^{35}$,
M.~Soares-Santos$^{36}$,
F.~Sobreira$^{37,19}$,
E.~Suchyta$^{38}$,
M.~E.~C.~Swanson$^{22}$,
G.~Tarle$^{25}$,
D.~Thomas$^{13}$,
V.~Vikram$^{39}$,
A.~R.~Walker$^{12}$
\begin{center} (DES Collaboration) \end{center}
}
\vspace{0.4cm}
\\
\parbox{\textwidth}{
Affiliations listed at the end of this document
}
}

\newcommand{\be}{\begin{equation}}
\newcommand{\ee}{\end{equation}}
\begin{document}
%\author[A. Salvador, F.J. S\'anchez et al.]
\maketitle
\date{\today}
\pagerange{--} \pubyear{2018}

%\label{firstpage}

\begin{abstract}
Non-linear bias measurements require a great level of control of potential systematic effects in galaxy redshift surveys. Our goal is to demonstrate the viability of using Counts-in-Cells (CiC), a statistical measure of the galaxy distribution, as a competitive method to determine linear and higher-order galaxy bias and assess clustering systematics. We measure the galaxy bias by comparing the first four moments of the galaxy density distribution with those of the dark matter distribution. We use data from the MICE simulation to evaluate the performance of this method, and subsequently perform measurements on the public Science Verification (SV) data from the Dark Energy Survey (DES). We find that the linear bias obtained with CiC is consistent with measurements of the bias performed using galaxy-galaxy clustering, galaxy-galaxy lensing, CMB lensing, and shear+clustering measurements. Furthermore, we compute the projected (2D) non-linear bias using the expansion $\delta_{g} = \sum_{k=0}^{3} (b_{k}/k!) \delta^{k}$, finding a non-zero value for $b_2$ at the $3\sigma$ level. We also check a non-local bias model and show that the linear bias measurements are robust to the addition of new parameters. We compare our 2D results to the 3D prediction and find compatibility in the large scale regime ($>30$ Mpc $h^{-1}$).

\end{abstract}


%--------------------------------------------------------------------------------------
%--------------------------------------------------------------------------------------
\section{Introduction}\label{intro}

	In recent years, photometric redshift galaxy surveys such as the Sloan Digital Sky Survey (SDSS) \citep{2017arXiv171103234K}, the Dark Energy Survey (DES) \citep{2016MNRAS.460.1270D}, and the future Large Synoptic Survey Telescope (LSST) \citep{2008arXiv0805.2366I} and Euclid \citep{2012SPIE.8442E..0ZA}, have arisen as powerful probes of the Large Scale Structure (LSS) of the universe and of dark energy. The main advantage of these surveys is their ability to retrieve information from a vast number of objects, yielding unprecedented statistics for different observables in the study of LSS. Their biggest drawback is the lack of line-of-sight precision and the systematic effects associated with it. Thus, well constrained systematic effects and robust observables are required in order to maximize the performance of such surveys. In this context, simple observables such as the galaxy number counts serve an important role in proving the robustness of a survey. In particular, the galaxy Counts-in-Cells (CiC), a method that consists of counting the number of galaxies in a given three-dimensional or angular aperture, has been shown to provide valuable information about the LSS~\citep{Peebles1980,1990MNRAS.247P..10E,Gaztanaga:1993ru,Bernardeau:1993qu,1998ApJ...497...16S} and gives an estimate of how different systematic effects can affect measurements. CiC can provide insights to higher-order statistical moments of the galaxy counts without requiring the computation resources of other methods \citep{Gil-Marin:2014sta}, such as the three- or four-point correlation functions. 

\noindent Understanding the relation between galaxies and matter (galaxy bias) is essential for the measurements of cosmological parameters \citep{Gaztanaga:2011yi}. The uncertainties in this relation strongly increase the errors in the dark energy equation of state or gravitational growth index \citep{Eriksen:2015hqa}. Thus, having a wide variety of complementary methods to determine galaxy biasing can help break degeneracies and improve the overall sensitivity for a given galaxy survey. 

\noindent In this paper we present a method to extract information from the galaxy CiC. Using this method, we measure the projected (angular) galaxy bias (linear and non-linear) in both simulations and observational data from DES, we compare the measured and predicted linear and non-linear bias, and we test for the presence of systematic effects. This dataset is ideal for this study since it has been already used for CiC in \citet{Clerkin:2016kyr}, where it was found that the galaxy density distribution and the weak lensing convergence ($\kappa_{WL}$) are well described by a lognormal distribution. The main difference between our study and \citet{Clerkin:2016kyr} is that our main goal is to provide a measurement of the galaxy bias, whereas \citet{Clerkin:2016kyr} study convergence maps.

\noindent \citet{2017arXiv171005045G} also perform CiC in DES data. Combining gravitational lensing information and CiC, they measure the galaxy density probability distribution function (PDF) and obtain cosmological constraints using the \texttt{redMaGiC} selected galaxies~\citep{2016MNRAS.461.1431R} in DES Y1A1 photometric data~\citep{2018ApJS..235...33D}. In our case we measure the moments of the galaxy density contrast PDF and compare them to the matter density contrast PDF from simulations (with the same redshift distributions) to study different biasing models, in a different galaxy sample (DES-SV).

\noindent Throughout the paper, we assume a fiducial flat $\Lambda$CDM+$\nu$ (one massive neutrino) cosmological model based on Planck 2013 + WMAP polarization + ACT/SPT + BAO, with parameters \citep{Ade:2013zuv} $\omega_b = 0.0222$, $\omega_c = 0.119$, $\omega_\nu = 0.00064$, $h = 0.678$, $\tau = 0.0952$, $n_s = 0.961$ and $A_s=2.21\times10^{-9}$ at a pivot scale $\overline{k}=0.05\rm{Mpc}^{-1}$ (yielding $\sigma_8 = 0.829$ at $z = 0$), where $h \equiv H_0 /100 \rm{km\ s}^{-1}\rm{Mpc}^{-1}$ and $\omega_i\equiv \Omega_i h^2$ for each species $i$.

\noindent 
The paper is organized as follows: in Section \ref{sec:data_sample}, we present the data sample used for our analysis. First, we present the simulations used to test and validate the method and afterwards, the dataset in which we perform our measurements. In Section \ref{sec:CiC}, we present the CiC theoretical framework and detail our method to obtain the linear and non-linear bias. Section \ref{sec:moments} and \ref{sec:DES-SV} present the CiC moments and bias calculations for the MICE simulation and DES-SV dataset, respectively. In Section \ref{sec:sys}, we study the systematic uncertainties in our method. Finally, in Section \ref{sec:conclusions}, we include some concluding remarks about this work.
%-----------------------------------------------------------------------------------------
%-----------------------------------------------------------------------------------------
\section{Data sample}\label{sec:data_sample}

%--------------------------------------------------------------------------------------------
%--------------------------------------------------------------------------------------------

\subsection{Simulations}\label{sec:sim}

In order to test and validate the methodology presented in this paper, we use the MICE simulation~\citep{2008MNRAS.391..435F, 2010MNRAS.403.1353C}. MICE is an N-body simulation with cosmological parameters following a flat $\Lambda$CDM model with $\Omega_m=0.25$, $\Omega_\Lambda=0.75$, $\Omega_b=0.044$, $n_s = 0.95$, and $\sigma_8 = 0.8$. The simulation covers an octant of the sky, with redshift z, between 0 and 1.4 and contains 55 million galaxies in the lightcone. The simulation has a comoving size $L_{box} = 3072h^{-1}$Mpc and more than $8\cdot10^9$ particles ~\citep{Crocce:2013vda}. The galaxies in the MICE simulation are selected following the procedure in \citet{2016MNRAS.455.4301C}, imposing the threshold $i_{evol}<22.5$. The MICE simulation has been extensively studied in the literature~\citep{2011MNRAS.411..277S,2016MNRAS.455.4301C,Hoffmann:2015mma,Pujol:2015wna,2018MNRAS.476.1071G}, including measurements of the higher-order moments in the dark matter field~\citep{2008MNRAS.391..435F}, providing an ideal validation sample.

%--------------------------------------------------------------------------------------------
%---------------------------------------------------------------------------------------------

\subsection{The DES SV Benchmark Data Sample}\label{sec:benchmark}

\begin{figure}
\includegraphics[width=0.45\textwidth]{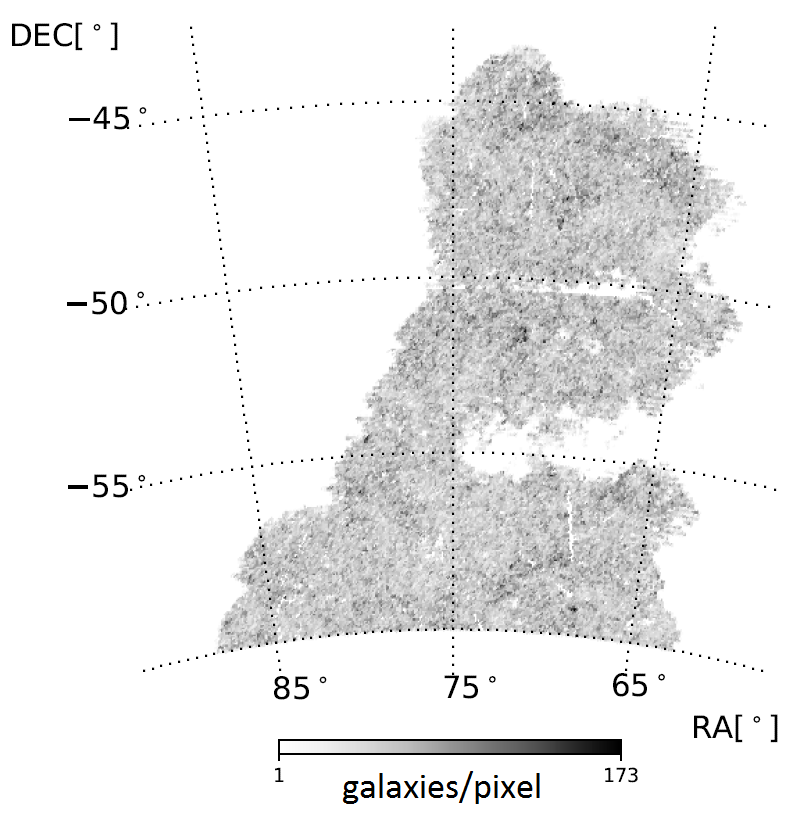}
\caption{Footprint of the DES SV benchmark sample~\citep{2016MNRAS.455.4301C}. We use approximately 2.3 million objects contained within this area for our studies.}
\label{fig:bench}
\end{figure}

\begin{figure}
\includegraphics[width=0.45\textwidth]{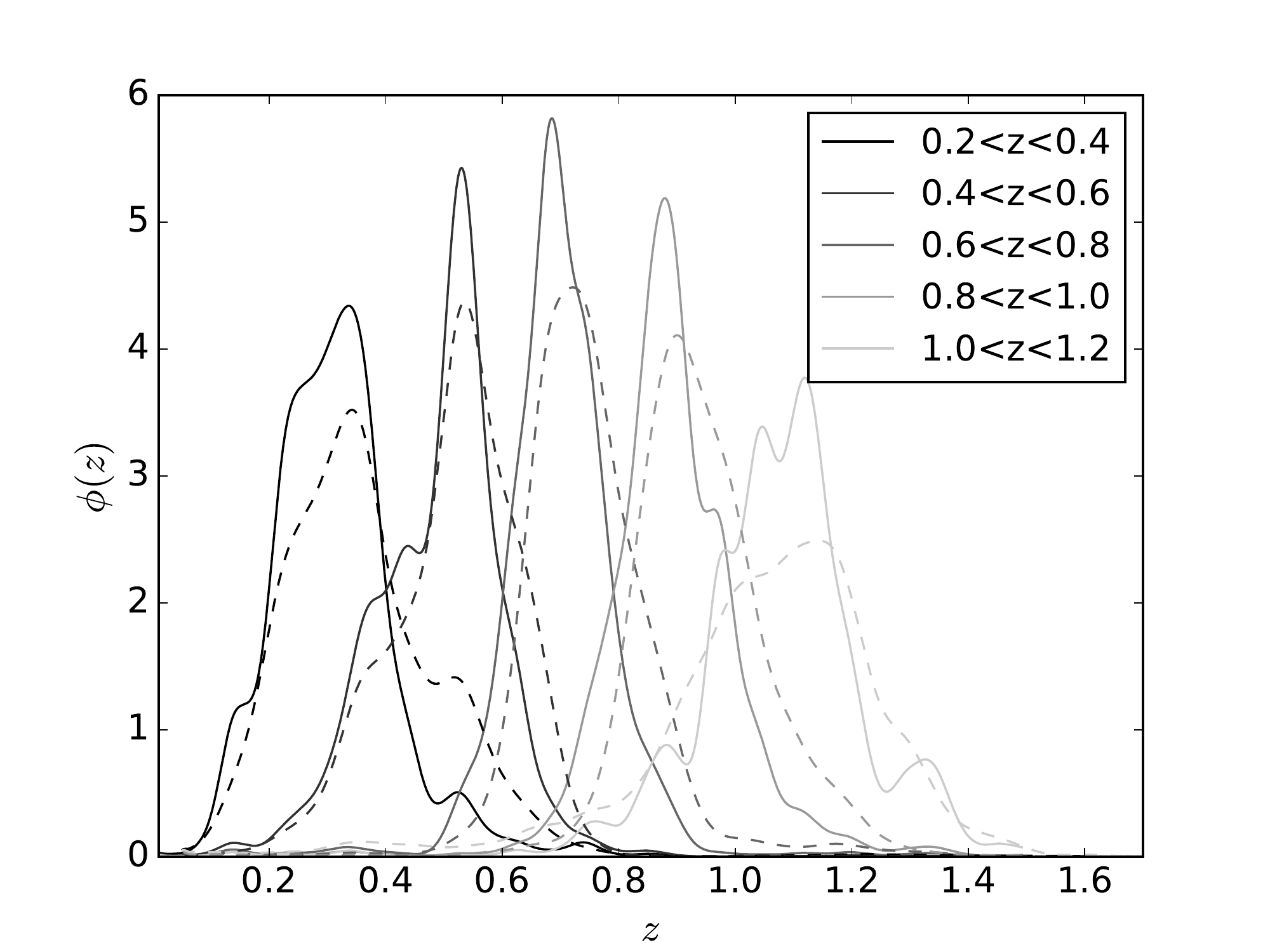}
\caption{Redshift distribution of the galaxies in each photometric redshift bin using TPZ (solid line) and BPZ (dashed line) in DES-SV benchmark data from \citet{2016MNRAS.455.4301C}. These distributions have been obtained by stacking the photometric redshift probability density functions of galaxies in a spectroscopic subsample detailed in ~\citet{2014MNRAS.445.1482S}. Lighter lines represent higher redshift slices.}
\label{fig:selection_function}
\end{figure}

In this paper we perform measurements of the density contrast distribution and its moments on the DES Science Verification (SV) photometric sample \footnote{This sample is available at \url{https://des.ncsa.illinois.edu/releases/sva1}} (Figure \ref{fig:bench}). The DES Science Verification observations were taken using DECam on the Blanco 4m Telescope near La Serena, Chile, covering over 250 deg$^{2}$ at close to DES nominal depth. From this sample we make selection cuts in order to recover the LSS Benchmark sample~\citep{2016MNRAS.455.4301C}. By doing this we minimize the possible two-point systematic effects and we ensure completeness. We focus on the SPT-E field, since it is the largest contiguous field and the best analyzed, with $60^{\circ}<$ RA $<95^{\circ}$, and $-60^{\circ} <$ Dec $< -40^{\circ}$ considering only objects with $ 18 < i < 22.5$ where $i$ is \texttt{MAG\_AUTO} as measured by \texttt{SExtractor}~\citep{Bertin1996} in the i-band. The star-galaxy separation is performed by selecting objects such that \texttt{WAVG\_SPREAD\_MODEL} $> 0.003$.  The total area considered for our study is then 116.2 deg$^{2}$ with approximately 2.3 million objects and a number density $n_{g} = 5.6$ arcmin$^{-2}$. Several photo-z estimates are available for these data~\citep{2014MNRAS.445.1482S}. We will focus on the TPZ \citep{2013MNRAS.432.1483C} and BPZ~\citep{Benitez:1998br} catalogs. We use the same 5 redshift bins used in \citet{2016MNRAS.455.4301C}. We use the redshift distributions from \citet{2014MNRAS.445.1482S}, which are depicted in Figure \ref{fig:selection_function}. These distributions have been obtained by comparing the DES-SV photometric sample including spectroscopic data from zCOSMOS~\citep{2007ApJS..172...70L,2009ApJS..184..218L} and VVDS Deep~\citep{2013A&A...559A..14L} among other datasets. For more details about the photometric redshift measurement and calibration, we refer the reader to~\citet{2014MNRAS.445.1482S}.

\noindent
Several measurements of the linear bias have been performed using this field~\citep{2016MNRAS.455.4301C,2016MNRAS.456.3213G,2018MNRAS.473.1667P}, making it ideal for this study.
%--------------------------------------------------------------------------------------------
%---------------------------------------------------------------------------------------------
\section{Theoretical framework and methodology}\label{sec:CiC}
\subsection{Counts-in-cells}
Counts-in-Cells~\citep{Peebles1980} is a method used to analyze the LSS based on dividing a galaxy survey into cells of equal volume ($V_{\rm{pix}}$) and counting the number of galaxies in each cell, ($N_{\rm{gal}}$). This method has also been extensively used in the literature to characterize the galaxy distribution~\citep{1990MNRAS.247P..10E,Gaztanaga:1993ru,Bernardeau:1993qu,1998ApJ...497...16S}, and, recently, even the neutral hydrogen in simulations~\citep{2018arXiv180809968L}. In the case of photometric redshift surveys, the lack of precision in the redshift determination makes angular aperture cells more appealing. Numerous examples of applications of CiC using angular aperture cells can be found in the literature~\citep{Gaztanaga:1993ru,2002ApJ...570...75S,Ross:2006zr,2011ApJ...729..123Y,2013MNRAS.435....2W}.

It is particularly useful to work with the density contrast, $\delta_{i}$ in each cell (or pixel), $i$, defined as:
\be\label{eq:delta}
\delta_{i}\equiv\frac{\rho_{i}}{\langle\rho\rangle}-1 
\ee
where $\rho_{i}\equiv\frac{N_{i,\rm{gal}}}{A_{i,\rm{pix}}}$ is galaxy density in the pixel of area $A_{i,\rm{pix}}$ and $\langle\rho\rangle$ is the mean density. In this work, we are going to use $\langle \cdots \rangle$ to denote statistical averages. Given these definitions, it follows that $\langle \delta \rangle = 0$.

In order to study the statistical properties of the density contrast distribution, $\delta$, we are interested in the measurement of the average of the $J$-point correlation functions, $\overline{w}_J(\theta)$, in a cell of solid angle $A = 2\pi(1-\cos{\theta})$~\citep{Gaztanaga:1993ru}:
\be \label{eq:Ipcfw}
\overline{w}_J(\theta) = \frac{1}{A^J}\int_{A}dA_1...dA_Jw_J(\theta_1,...,\theta_J), \, J \geq 2
\ee
with $dA_{i}=\sin{\theta_{i}}d\theta_{i}d\phi_{i}$ and $w_{J}(\theta)$ the $J$-point angular correlation function. 

To estimate the angle-averaged $J$-point correlation function, $\overline{w}_{J}(\theta)$, we use the \textbf{corrected connected moments}, $\langle \delta^{J} \rangle_{c}$, taking into account the discrete nature of CiC and assuming Poisson-like shot-noise contributions as introduced by~\citet{Gaztanaga:1993ru}. In particular, we are interested in terms up to $J=4$:
\be
\begin{gathered}
\overline{w}_{2}(\theta)=\langle\delta^2\rangle_c=\langle\delta^2\rangle-\frac{1}{\overline{N}}\\
\overline{w}_{3}(\theta)=\langle\delta^3\rangle_c=\langle\delta^3\rangle-\frac{3}{\overline{N}}\langle\delta^2\rangle_c-\frac{1}{\overline{N}^2}\\
\overline{w}_{4}(\theta)=\langle\delta^4\rangle_c=\langle\delta^4\rangle-3\langle\delta^2\rangle^2-\frac{7}{\overline{N}^2}\langle\delta^2\rangle_c-\frac{6}{\overline{N}}\langle\delta^3\rangle_c-\frac{1}{\overline{N}^3}
\end{gathered}
\ee

where $\overline{N}=\frac{N_{\rm{gal}}^{tot}*A_{\rm{pix}}}{A_{\rm{tot}}}$, and $N_{\rm{gal}}^{tot}$ the total number of galaxies, $A_{\rm{tot}}$ the total area, and $A_{\rm{pix}}$ the area of the pixel.

\begin{figure}
\includegraphics[scale=0.45]{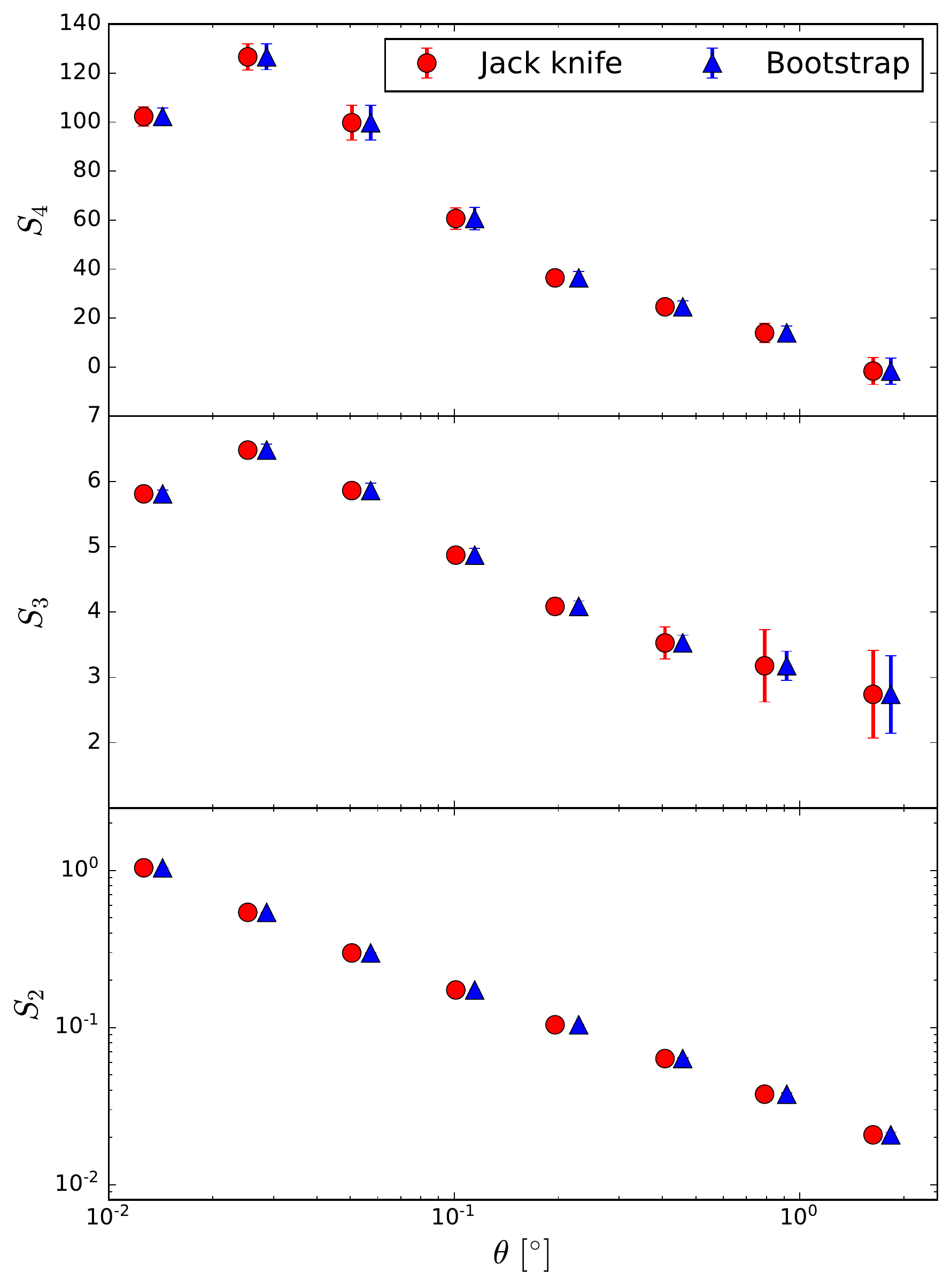}
\caption{Moments of the density contrast distribution as a function of cell scale in the MICE simulation for the redshift bin $0.2<z<0.4$, with jackknife errors (solid circles) and bootstrap errors (solid triangles). The results for a given scale $\theta$ have been separated in the figure for visualization purposes, being the blue triangles the ones shown at the nominal measured scale.}
\label{fig:errors}
\end{figure}

For our study we use the \textbf{rescaled connected moments} $S_J$ defined as:
\be\label{eq:hierarchyw}
S_J\equiv\frac{\overline{w}_J(\theta)}{[\overline{w}_2(\theta)]^{J-1}}, \hspace{4pt} J>2
\ee
\be
S_{2} = \overline{w}_{2}(\theta)
\ee
In most previous studies, the cells considered were spheres with radii of varying apertures (\citet{Peebles1980}, \citet{Bernardeau:1993qu}). We perform our measurements of the projected (angular) density contrast by dividing the celestial sphere into \texttt{HEALpix} pixels \citep{Gorski2005}. 
\noindent
For our study we vary the \texttt{HEALPix} parameter $N_{side}$ from 32 to 4096 (\textit{i.e.} apertures ranging from $1.83^\circ$ to $0.014^\circ$). The angular aperture, $\theta$, is estimated as the square root of the pixel area. According to equation (\ref{eq:Ipcfw}) there is a dependence on the boundaries of the cell and thus on the shape that we choose for the pixels. \citet{Gaztanaga:1993ru}, estimates CiC for square cells of side $l$ in a range $l = 0.03^\circ-20 ^\circ$ and compares to the average correlation functions $\overline{w}_2(\theta)$. The agreement between the two estimates indicates that square cells give very similar results to circular cells when the sizes of the cells are scaled to $\theta=l/\sqrt{\pi}$. Using data from MICE, we perform several tests to see that the concrete shape of the pixel, when it is close to a regular polygon, does not affect the measured moments despite boundary effects (Appendix \ref{sec:appendix_b}). Furthermore, when working with the acquired observational data, the geometry of the survey becomes complicated. A discussion of how we deal with this is found in Appendix \ref{sec:appendix_c}. 
\noindent
The error bars throughout this paper are estimated using the bootstrap method~\citep{Efron79,2006ASPC..357..271M,astroMLText}. This choice is mainly due to the lack of number of samples for large pixel sizes that might limit the precision of other methods such as jack-knife, given that the latter depends highly on the number of samples as pointed out in \citep{2009MNRAS.396...19N}. Figure \ref{fig:errors} shows agreement between the uncertainties computed using the jackknife and bootstrap methods for a randomly chosen redshift bin in the MICE simulation. We use $N_{b}=\min{\left(N_{pix},100\right)}$ bootstrap and jack-knife realizations of the density contrast distribution to estimate our errors, where $N_{pix}$ is the total number of unmasked pixels in our map. 

%--------------------------------------------------------------------------------------
\subsection{Galaxy bias}\label{sec:bias}

One of the most important applications of the CiC observable is the determination of the galaxy bias. We observe the galaxy distribution and use it as a proxy to the underlying matter distribution. Both baryons and dark matter structures grow around primordial overdensities via gravitational interaction, so these distributions should be highly correlated. This relationship is called the \textit{galaxy bias}, which measures how well galaxies trace the dark matter.
\noindent
Galaxy biasing was seen for the first time analyzing the clustering of different populations of galaxies \citep{1978ApJ...221....1D,1980ApJ...236..351D}. The theoretical relation between galaxy and mass distributions was suggested by \citet{Kaiser:1984sw} and developed by \citet{1986ApJ...304...15B}. Since then, many different prescriptions have arisen \citep{Fry:1992vr,Bernardeau:1995ty,Mo:1995cs,Sheth:1999mn,Manera:2009ak,2011MNRAS.415..383M}. However, there is no generally accepted framework for galaxy biasing. While the galaxy and dark matter distribution are related, the exact relation depends on galaxy formation \citep{PressSchechter}, galaxy evolution \citep{Nusser:1993sx,Tegmark:1998wm,2000ApJ...531....1B}, and selection effects. Bias depends strongly on the environment. Using dark matter simulations, \citet{Pujol:2015wna} show how the halo bias is determined by local density and not by halo mass. Several studies have demonstrated the different behaviors of early-type and late-type galaxies at both small and large scales 
\citep{Ross:2006zr,Willmer:1999hf,2002ApJ...571..172Z,2002MNRAS.332..827N}. To have a good estimate of the real matter distribution, it is convenient to use a galaxy sample as homogeneous as possible. 
With the linear bias $b(z)$ approximation, we can relate the matter fluctuations $\delta_m$ with the fluctuations in the galaxy distribution $\delta_g$:
\be 
\delta_g=b\delta_m
\ee
In the linear approximation, up to scalings, all statistical properties are preserved by the biasing and the observed galaxy properties reflect the matter distribution on large scales, as long as we consider only two-point statistics. However, in the general case, it is highly unlikely that the relation is both local and linear. Non-local dependencies might come from some properties such as the local velocity field or derivatives of the local gravitational potential \citep{Fry:1992vr,Scherrer:1997hp}. Bias also depends on redshift \citep{Fryevolution,Tegmark:1998wm}.
\noindent
When non-Gaussianities are taken into account, linear bias fails to be a good description. If we want to measure higher orders we can assume that the (smoothed) galaxy density can be written as a function of the mass density and expand it as a Taylor series (assuming a local relation) \citep{friemannGaz,Fry:1992vr}:
\be\label{eq:exp}
\delta_g=f(\delta)=\sum_{k=0}^\infty\frac{b_k}{k!}\delta^k_m
\ee
The linear term $b_1=b$ is the usual linear bias. Using this expansion we can relate the dark matter and the galaxy density contrast moments using the following relationships~\citep{Fry:1992vr}:
\begin{equation}
S_{2, mod}=b^{2}S_{2m}
\label{eq:S2_S2m}
\end{equation}
\begin{equation}
S_{3, mod}=b^{-1}(S_{3m}+3c_2)
\label{eq:S3_S3m}
\end{equation}
\begin{equation}
S_{4, mod}=b^{-2}(S_{4m}+12c_2S_{3m}+4c_3+12c_2^2)
\label{eq:S4_S4m}
\end{equation}
where $c_k=b_k/b$ for $k\geq2$, the subscript $m$ refers to the underlying matter distribution, and the subscript $mod$ to the galaxy distribution. We will refer to this model as \textbf{local}.

\citet{Bel:2015jla} point out that ignoring the contribution from the non-local bias can affect the linear and non-linear bias results. As a consequence, we analyze the case when the non-local contribution is included. To do so, we substitute $c_{2}$ by $c'_{2}=c_{2}-\frac{2}{3}\gamma_{2}$, where $\gamma_{2}$ is the so-called non-local bias parameter \citep{Bel:2015jla}. We will refer to this model as \textbf{non-local}.

Note that we omit the terms higher than 3rd order because, as we will show later, we have very limited sensitivity to $b_{3}$, and expect to have no sensitivity to $b_{4}$. 

%Finally, the authors in \citet{2017arXiv171005045G} find that the data prefers the inclusion of a stochastic component. Therefore, we consider a model (described in Appendix \ref{app:stochastic_bias}) which we refer to as \textbf{local+stochastic} if the stochastic contribution is considered but the non-local contribution is ignored and \textbf{non-local+stochastic} if the stochastic and the non-local contributions are considered.
\subsection{Estimating the projected linear and non-linear bias}
\begin{figure}
\centering
\includegraphics[width=0.45\textwidth]{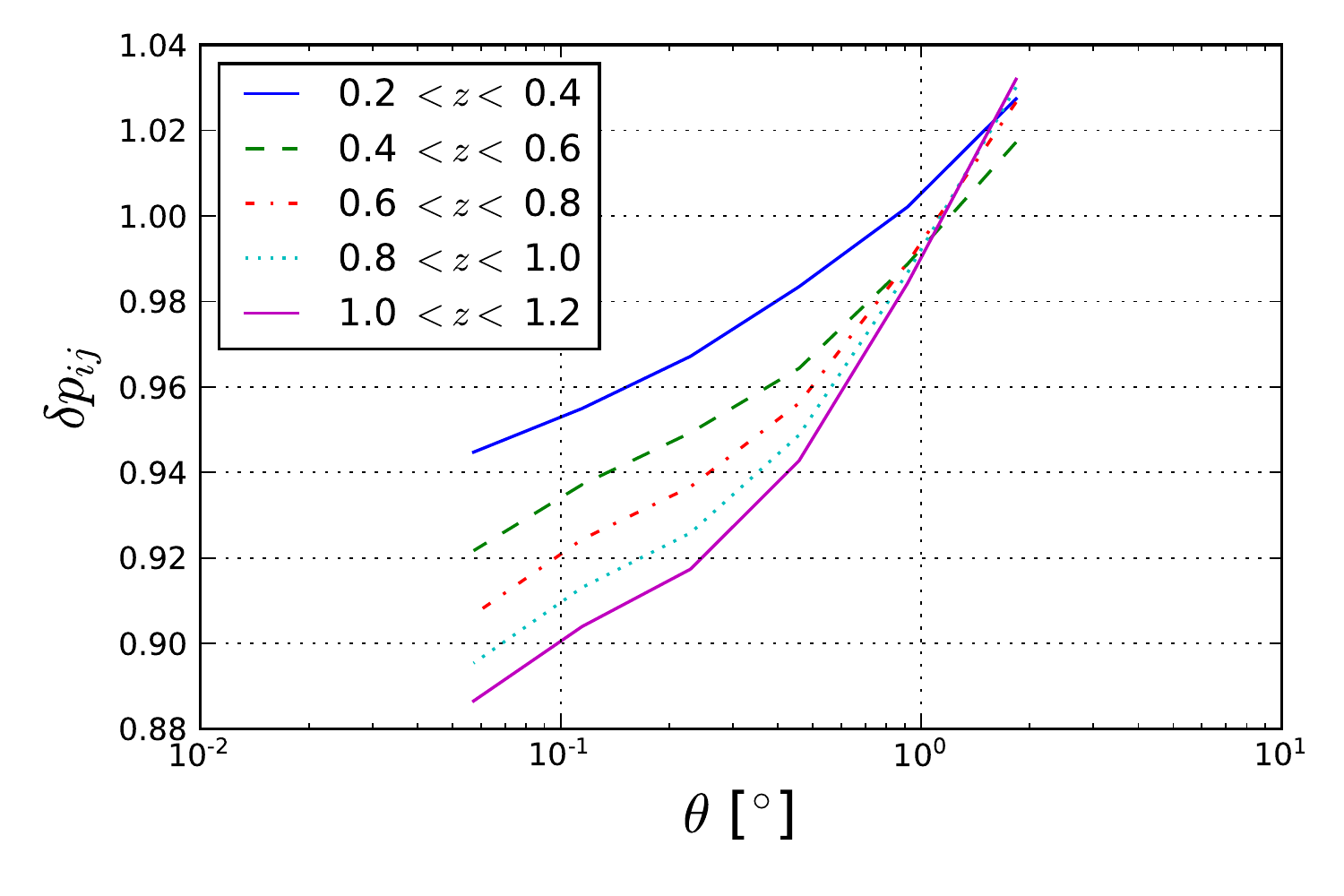}
\caption{Ratio between $S_{2,m}$ for $\Omega_{m}=0.2$ (the minimum allowed by Planck priors) and $S_{2,m}$ for our fiducial model as a function of the cell aperture angle $\theta$. The different lines represent different redshift bins. We see that the variation is within 12\% of the linear prediction.}
\label{fig:variation_S2}
\end{figure}

The relations in equations (\ref{eq:S2_S2m}-\ref{eq:S4_S4m}) refer to the three-dimensional case and connect an observed galaxy distribution with its underlying dark matter distribution, both tracing the same redshift range and cosmological parameters. We assume that this bias model is also valid for the projected moments (we will check the validity of this assumption later). Moreover, given the measurements in a dark matter simulation with the same redshift distribution and angular footprint as our galaxy dataset, we estimate the linear and non-linear bias of these galaxies using equations (\ref{eq:S2_S2m}-\ref{eq:S4_S4m}). Note that these relations apply when we are comparing two datasets with the same value for $\sigma_{8}$ parameter. In the case that $\sigma_{8} \neq \sigma_{8,m}$ we will have to correct the resulting bias so,
\begin{equation}
b_{corr} = b_{uncorr}\frac{\sigma_{8,m}}{\sigma_{8}}.
\label{eq:bias_corr}
\end{equation}
We will use this correction in Section~\ref{ssec_sys_s8}. We also take advantage of the fact that the skewness and kurtosis depend weakly with the cosmological parameters~\citep{Bouchet:1992uh}. In particular, a 5\% variation choosing $\Omega_{m}=0.25$ translates to a variation of $0.2\%$ in the measured $S_{3m}$, which is much smaller than the statistical fluctuations that we expect from our samples. In the case of $S_{4m}$ our sensitivity is even lower, making it safe to use a simulation with the same footprint and redshift distribution, as long as the variation in the cosmological parameters is small. However, this is not necessarily true for the case of $S_{2m}$, where the dependency on the cosmological parameters is higher. We check this using equation (\ref{eq:Ipcfw}) to compute the projected $S_{2m}$ for two different sets of cosmological parameters: our fiducial Planck cosmology~\citep{Ade:2013zuv} and a model with $\Omega_{m}=0.2$. We use a Gaussian selection function $\phi(z)$ with $\sigma_{z}=0.05(1+z)$ since this is representative of the datasets that we analyze in this work. After this, we check the ratio:
\begin{equation}
\delta p_{ij}(z,\theta) = \frac{S_{2m,i}(z,\theta)}{S_{2m,j}(z,\theta)}\frac{D^{2}_{+,j}(\bar{z})}{D^{2}_{+,i}(\bar{z})}
\end{equation}
for the different redshift slices considered in our analysis, where the subscripts $i$ and $j$ correspond to two different sets of cosmological parameters and $D_{+}(\bar{z})$ is the linear growth factor \citep{Peebles1980,Heath1977} evaluated at the mean redshift of the considered slice. This gives us an upper limit to the expected variation in $S_{2m}$ to consider in our analysis. In Figure~\ref{fig:variation_S2} we can see that the variation is within 12\% of the linear prediction, thus, we conservatively assign 12\% systematic error to $S_{2m}$ due to this variation.
Under these conditions we perform a simultaneous fit to $b$, $b_{2}$, $b_{3}$ and $\gamma_{2}$. In order to do so, we consider the likelihood:
\begin{multline}
\log{\mathcal{L}} = -\frac{1}{2}\sum_{k=2}^{4}\sum_{i,j}\left[S_{k,g}(\theta_{i})-S_{k,mod}(\theta_{i})\right]\\
C^{-1}_{k,ij}\left[S_{k,g}(\theta_{j})
-S_{k,mod}(\theta_{j})\right] = -\frac{\chi^{2}}{2}
\label{eq:loglike}
\end{multline}
where $S_{k,g}$ are the measured galaxy moments and $S_{k,mod}$ are the models in equations (\ref{eq:S2_S2m}), (\ref{eq:S3_S3m}), and (\ref{eq:S4_S4m}). We checked that the measured $S_{k}$ follow a Gaussian distribution. The covariances $C_{k,ij}$ are computed as follows:
\begin{equation}
C_{k,ij}=\frac{N_{u,pix}(\theta_{i})}{N_{u,pix}(\theta_{j})}2^{2(j-i)}\sigma_{k}(\theta_{i})\sigma_{k}(\theta_{j})
\end{equation}
with $N_{u,pix}(\theta_{i})$ being the number of pixels used in an aperture, $\theta_{i}$. Note that, since we are using \texttt{HEALPix}, which imposes a fixed grid, and we are not repeating the measurements in translated/rotated galaxy fields, we are re-using the same galaxies for different scales, so the factor $\frac{N_{u,pix}(\theta_{i})}{N_{u,pix}(\theta_{j})}2^{2(j-i)}$ accounts for the induced correlation due this reuse. We assume that the errors in the dark matter moments and the errors in the galaxy moments are not correlated and add them in quadrature, so: \begin{equation}
\sigma_{k}(\theta_{i})=\sqrt{\sigma_{k,g}^{2}(\theta_{i})+\sigma_{k,m}^{2}(\theta_{i})}
\end{equation}
where $\sigma_{k,g/m}(\theta_{i})$ is the standard deviation of the $k$-th (galaxy or matter) moment in an aperture $\theta_{i}$ computed using bootstrapping.

We use the following flat priors:
\begin{itemize}
\item $0 < b < 10$.
\item $-10 < b_{2} < 10$.
\item $-10 < b_{3} < 10$.
\item $\gamma_{2}=0$ (or in the case of non-local model $-10 < \gamma_{2} < 10$).
\end{itemize}
These priors have been chosen to prevent unphysical results. We evaluate the likelihood and obtain the best fit values and their uncertainties by performing a MCMC using the software package \texttt{emcee}~\citep{2013PASP..125..306F}. Summarizing, the method works as follows:
\begin{enumerate}
\item Measure CiC moments using \texttt{HEALPix} pixels in the galaxy sample.
\item Measure CiC moments using the same pixels and selection function in a dark matter simulation with comparable cosmological parameters.
\item Evaluate statistical and systematic uncertainties in the measured moments.
\item Obtain best fit $b, b_{2}, b_{3}$, (and $\gamma_{2}$ in the non-local model) using MCMC with the models from equations (\ref{eq:S2_S2m}-\ref{eq:S4_S4m}).
\end{enumerate}

In summary, in the local model we fit 3 free parameters, whereas in the non-local model we fit 4.

\citet{Hoffmann:2015mma} present a prediction for the non-linear bias as a function of the linear bias in the three-dimensional case: 
\begin{eqnarray}
\centering
b_{2}&=&b^{2}-2.45 b+1.03\\
b_{3}&=&b^{3}-7.32 b^{2}+10.79
b-3.90
\label{eq:hoffman}
\end{eqnarray}
We will use these predictions to test the compatibility between the 3D and the measured projected values for the non-linear bias.
%--------------------------------------------------------------------------------------

%--------------------------------------------------------------------------------------
%--------------------------------------------------------------------------------------
\section{RESULTS IN SIMULATIONS} \label{sec:moments}

In order to validate this method, we first compute the CiC moments in the MICE simulation (in both galaxies and DM) using a Gaussian selection function $\phi(z)$ with $\sigma_z=0.05(1+z)$. This $\sigma_z$ is similar to the photometric redshifts found in the data using TPZ ~\citep{Kind:2013eka} and BPZ ~\citep{Benitez:1998br}. We split our sample into 5 photometric redshift bins: $z \in [0.2, 0.4], [0.4, 0.6], [0.6, 0.8], [0.8, 1.0], [1.0, 1.2]$, mirroring the choice in \citep{2016MNRAS.455.4301C}. Then we do the same with the SV data sample presented in Section \ref{sec:benchmark} with TPZ photometric redshifts.

\subsection{Angular moments for MICE}\label{sec:mice_cic}

\begin{figure}
\includegraphics[scale=0.45]{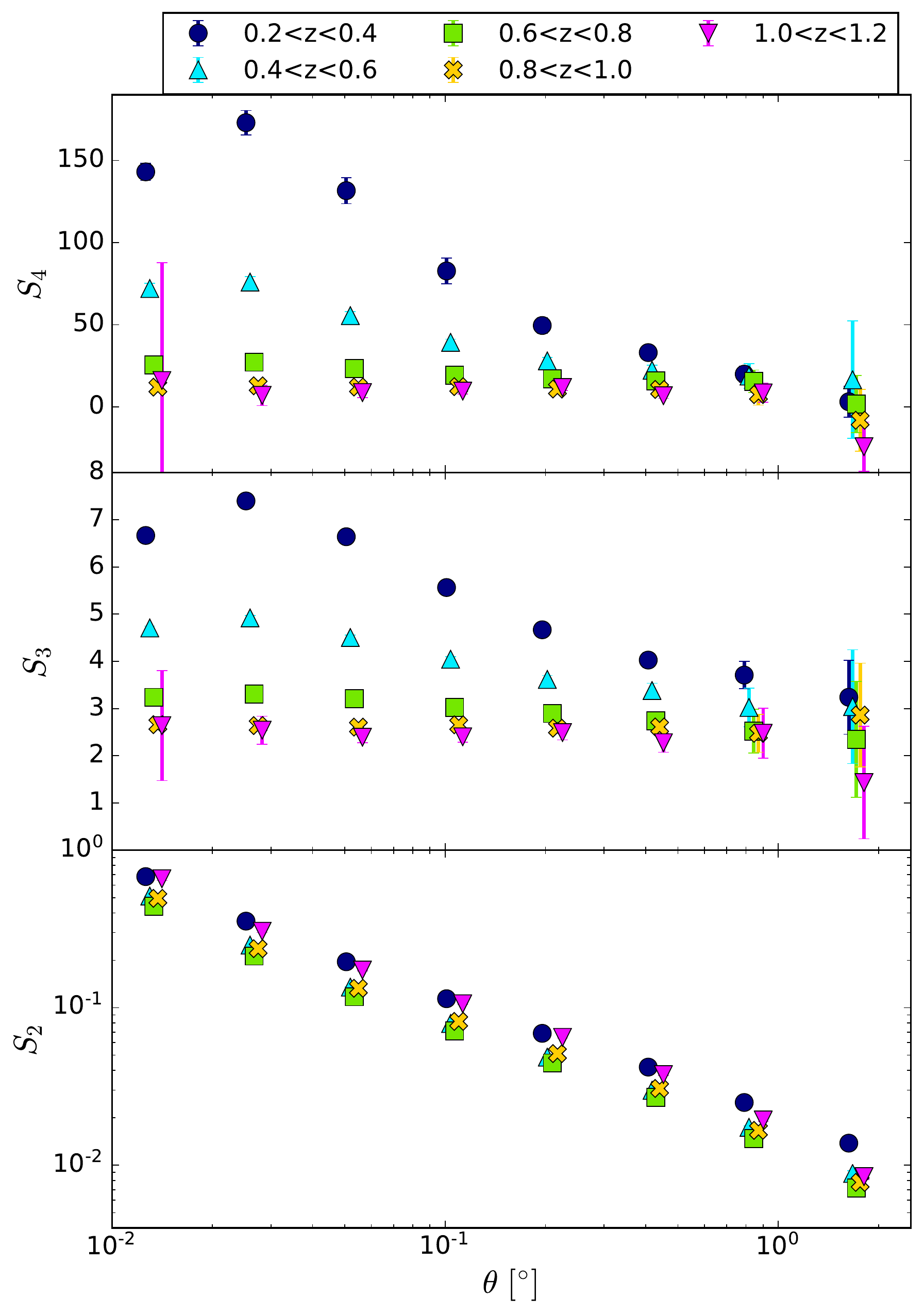}
\caption{Moments of the density contrast distribution as a function of cell scale in the MICE simulation with Gaussian photometric redshift ($\Delta z=0.2$ $\sigma_z=0.05(1+z)$) for different redshift bins. The results for a given scale $\theta$ have been separated in the figure for visualization purposes.}
\label{fig:micemomphotoz}
\end{figure}

Figure \ref{fig:micemomphotoz} shows the moments of the density contrast distribution as a function of the cell scale for the different photometric redshift bins. We observe that the moments follow the expected trend, that is, lower redshift bins have higher values for the higher-order moments since non-linear gravitational collapse has a larger effect on these. This is true for all measurements except for the last two redshift bins of the variance, $S_{2}$. This can be due to the magnitude cuts, since the galaxy populations are different at different redshifts. We also see that the larger the cell scale, the smaller the variance $S_{2}$, since larger cell scales should be more homogeneous. The skewness and the kurtosis in linear scales ($\theta>0.1^\circ$) are constant and of the same order of magnitude as the expected values ($S_{3} \approx 34/7$, $S_{4} \approx 60712/1323$) ~\citep{Bernardeau:1993qu}. The behavior at non-linear scales is due to the non-linearities of the MICE simulation.

%--------------------------------------------------------------------------------------
%--------------------------------------------------------------------------------------
%--------------------------------------------------------------------------------------
%--------------------------------------------------------------------------------------
\subsection{Projected galaxy bias in MICE simulation}\label{sec:mice_nlresults}

\begin{figure}
\centering
\includegraphics[scale=0.45,trim={0 2.5cm 0 2cm}, clip]{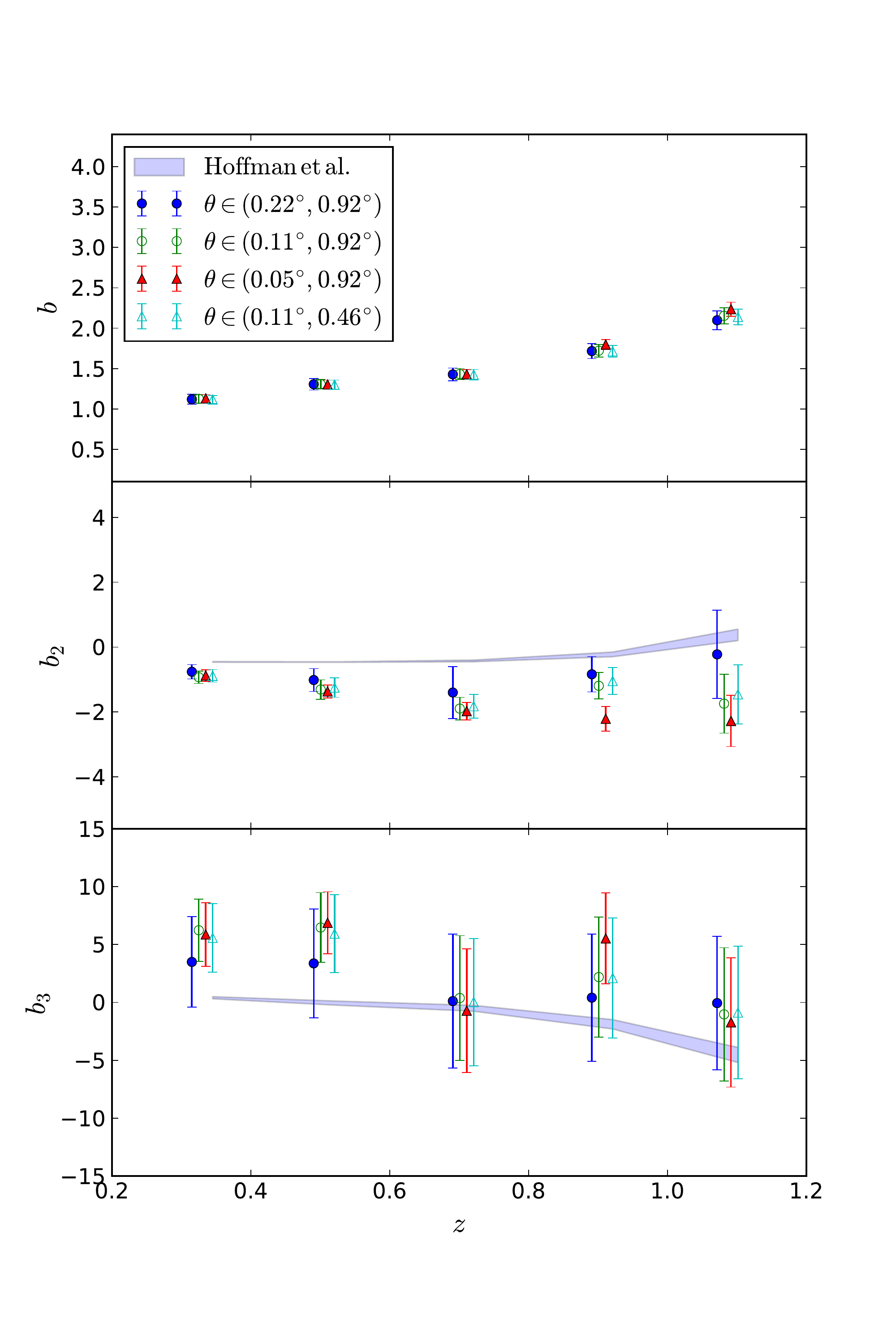}
\caption{Linear and non-linear bias results as a function of redshift for MICE data with Gaussian photo-z. The different marker shapes represent the best-fit results considering different ranges of the aperture angle $\theta$. For the solid triangles we consider the range from $0.05^{\circ}$ to $0.92^{\circ}$, open circles are our fiducial case with $0.11^{\circ} < \theta < 0.92^{\circ}$, for the solid circles, we take out the smallest scale in our fiducial case and in open triangles we take out the largest scale. The top panel shows the projected linear bias $b$ as a function of redshift, the middle panel shows the best-fit results for the projected $b_{2}$, and $b_{3}$ is shown in the lower panel. The shaded region corresponds to the 3D predicted values using equations (\ref{eq:hoffman}). The results for a given redshift $z$ have been separated in the figure for visualization purposes.}
\label{fig:nl_bias_summary_mice}
\end{figure}

We smear the true redshift with the proper selection function in the MICE dark matter field, obtained from a dilution of the dark matter particles (taking $1/700$ of the particles). \citet{2016MNRAS.459.3203C} demonstrate that the dilution of the dark matter field does not impact their statistics and using the measured moments from the previous section we proceed to perform a simultaneous fit for $b$, $b_{2}$, and $b_{3}$ using the local, non-linear bias model from equations (11,12,13). The fit results are summarized in Figure \ref{fig:nl_bias_summary_mice}. We can see the impact of changing the range of $\theta$ considered in the fit. In this case we see that including scales smaller than $0.1^{\circ}$, where non-linear clustering has a large impact, affects the $b_{2}$ results. This, together with the fact that the reduced $\chi^{2}$ minimum value doubles when including $\theta = 0.05^{\circ}$ clearly shows that we should not consider scales smaller than $\theta = 0.1^{\circ}$. We can see as well that $b_{3}$ is compatible with zero and that we have a limited sensitivity to it, given the area used. Thus, the choice of ignoring terms of orders higher than $b_{3}$ becomes a good approximation. However, for $b_{2}$ we are able to measure a significant non-zero contribution. We can also see that the predicted values for the 3D non-linear bias parameter $b_{2}$ are not in good agreement at small scales, while there is an indication of better agreement at larger scales. This suggests that the 2D and 3D values for $b_{2}$ might be compatible at larger scales, in agreement with \citet{2011MNRAS.415..383M} who show that the local bias is consistent for scales larger than $R>30-60$ Mpc/$h$. They also show that the values of $b_1$ and $b_2$ vary with the scale and converge to a constant value around $R>30-60$ Mpc/$h$, which means that the values that we measure here have not yet fully converged. The prediction for $b_{3}$ seems to be compatible with the estimated values given the size of the error bars. These results show that we should consider $b_{2}$ as a first order (small) correction to the linear bias model at these scales for projected (angular) measurements.
The individual fits can be seen in Appendix \ref{app:fits}.
\subsection{Verification and biasing model comparison}
\begin{figure}
\centering
\includegraphics[scale=0.45,trim={0 1.6cm 0 2cm}, clip]{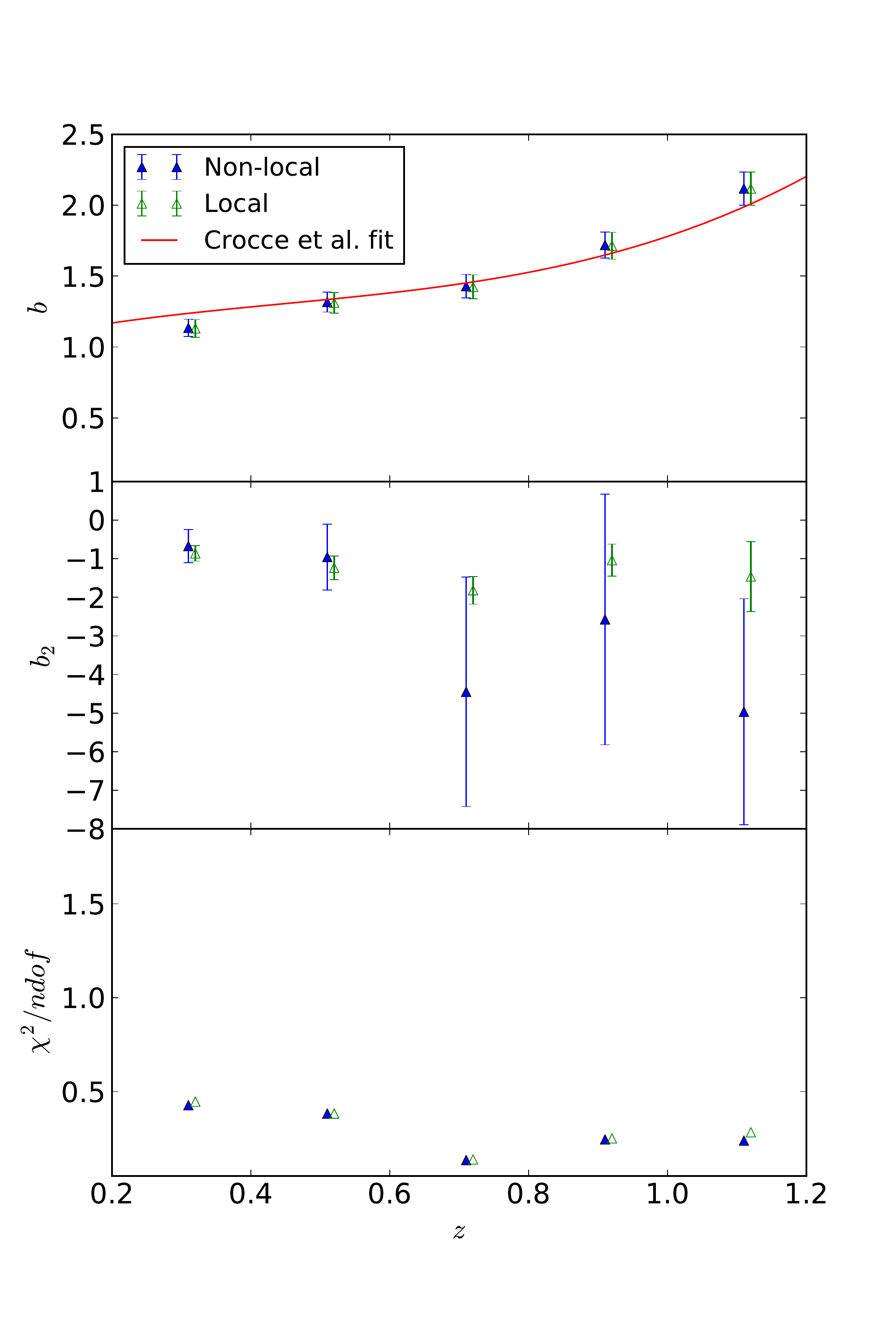}
\caption{(Top) Comparison between the MICE simulation bias obtained using CiC with different biasing models: non-local (solid triangles), and local (open triangles). We also show the best-fit from \citet{2016MNRAS.455.4301C} (Figure 17) as reference. The middle panel shows the equivalent results for $b_{2}$. This is done for Gaussian photo-z with $\sigma_{z}=0.05(1+z)$. (Bottom) total reduced chi-square for each of the models when fitting the moments to obtain the bias.}
\label{fig:linear_bias_mice}
\end{figure}

In order to verify this method and check if the local non-linear model considered induces certain systematic biases on the results, we check that the measured linear bias is compatible with corresponding measurements from the two point correlation function (Figure \ref{fig:linear_bias_mice}). In particular, we use the best fit parametrization from \citet{2016MNRAS.455.4301C}:
\begin{equation}
b_{best}(z) = 0.98 + 1.24z - 1.72z^{2} + 1.28z^{3}
\end{equation}
In Figure~\ref{fig:linear_bias_mice}, we can see that the local and non-local bias are in agreement, most likely due to the scale range that we are dealing with and the projection effects due to the size of the redshift slices. In this figure, we can also notice that the reduced chi-square for both models is similar, and that they are well below one. Given that the number of degrees of freedom is small, it is still possible that these values are correct, however, it is unlikely that this happens for all redshift bins. This suggests that, in agreement with \citet{2009MNRAS.396...19N}, bootstrapping uncertainties are overestimated. However, we prefer to use these conservative uncertainties, rather than state uncertainties that are too optimistic since, one of the main goals of this work will be to state the statistical significance on the non-linear $b_{2}$ term. Another interesting feature in Figure~\ref{fig:linear_bias_mice} is that the uncertainties in $b_{2}$ for the non-local model are considerably larger than in the local model. This is due to the fact that $\gamma_{2}$ is highly correlated with $b_{2}$, which makes the posterior distribution for $b_{2}$ much wider, increasing the resulting uncertainty.

\section{RESULTS IN DES-SV data}
\label{sec:DES-SV}
\subsection{Angular moments for DES - SV}\label{sec:benchcic}
\begin{figure}
\includegraphics[scale=0.45]{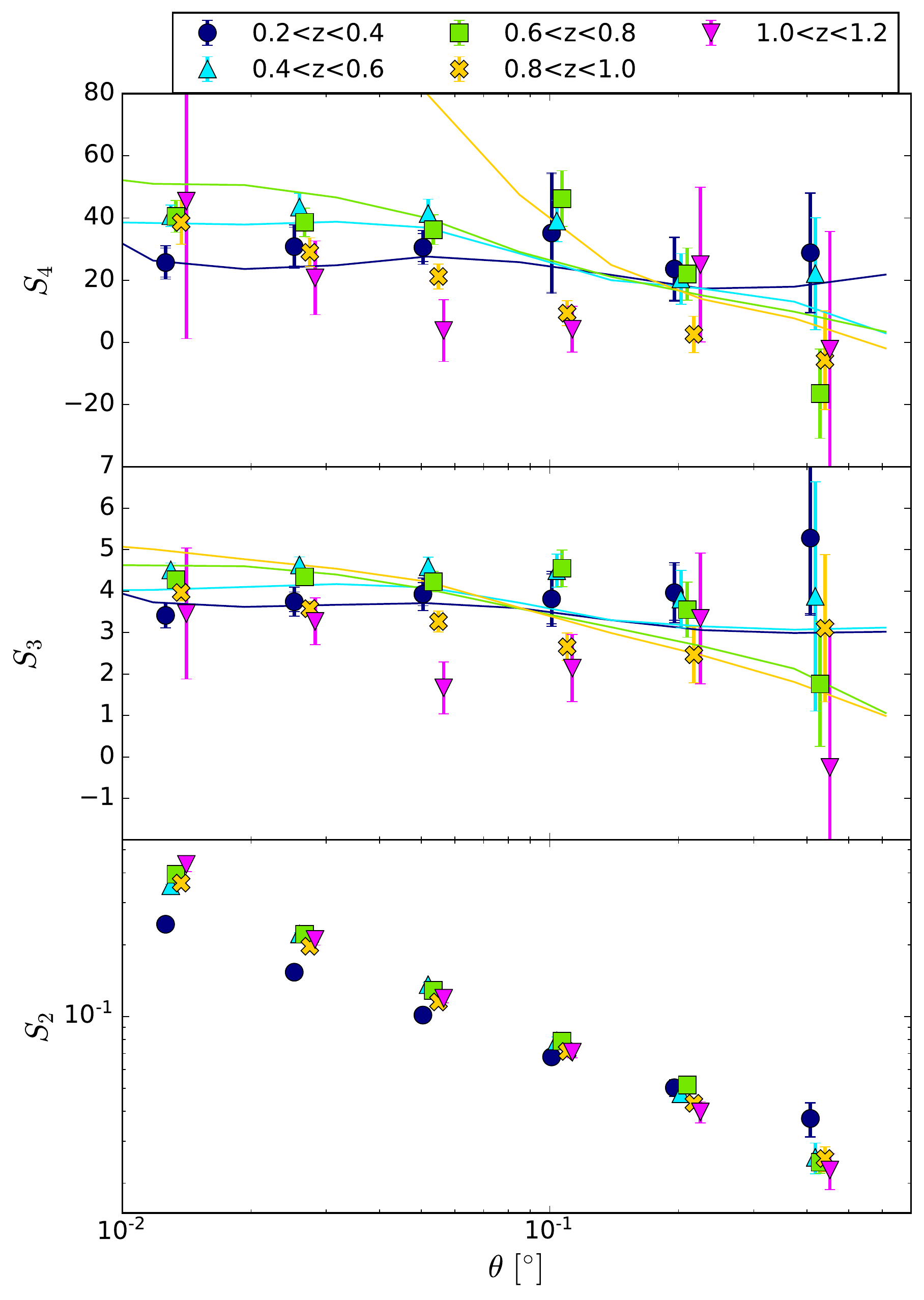}
\caption{Moments of the density contrast distribution of the DES SV benchmark sample as a function of cell scale, for five different redshift bins and different scales. The results for a given scale $\theta$ have been separated in the figure for visualization purposes. We compare with the results from \citet{2013MNRAS.435....2W} for CFHTLS marked with solid lines of different colors for the different redshift bins: navy $(0.2 < z < 0.4)$, cyan $(0.4 < z < 0.6)$, lime $(0.6 < z < 0.8)$, yellow $(0.8 < z < 1.0)$.}
\label{fig:benchmom}
\end{figure}

Using the same footprint, selection cuts, and redshift bins as in \citet{2016MNRAS.455.4301C}, we compute the moments of the density contrast distribution for the SV data. These results are depicted in Figure \ref{fig:benchmom} as a function of cell scale for different redshift bins. Here, as in the case of MICE, the variance decreases with the scale. The skewness and the kurtosis are also constant and of the same order of magnitude as the theoretical values within errors. The largest differences when compared with the simulation are in the non-linear regime due to the different way non-linearities are induced in the simulation and in real data. We also compare to the results from CFHTLS found in \citet{2013MNRAS.435....2W}. We find a similar general behavior, as well as the same order of magnitude in the measured $S_{3}$ and $S_{4}$. However, we do not expect the same exact results since the redshift distributions from CFHTLS do not match exactly the corresponding distributions in the DES-SV data.

\begin{figure}
\centering
\includegraphics[scale=0.45,trim={0 2.5cm 0 2cm}, clip]{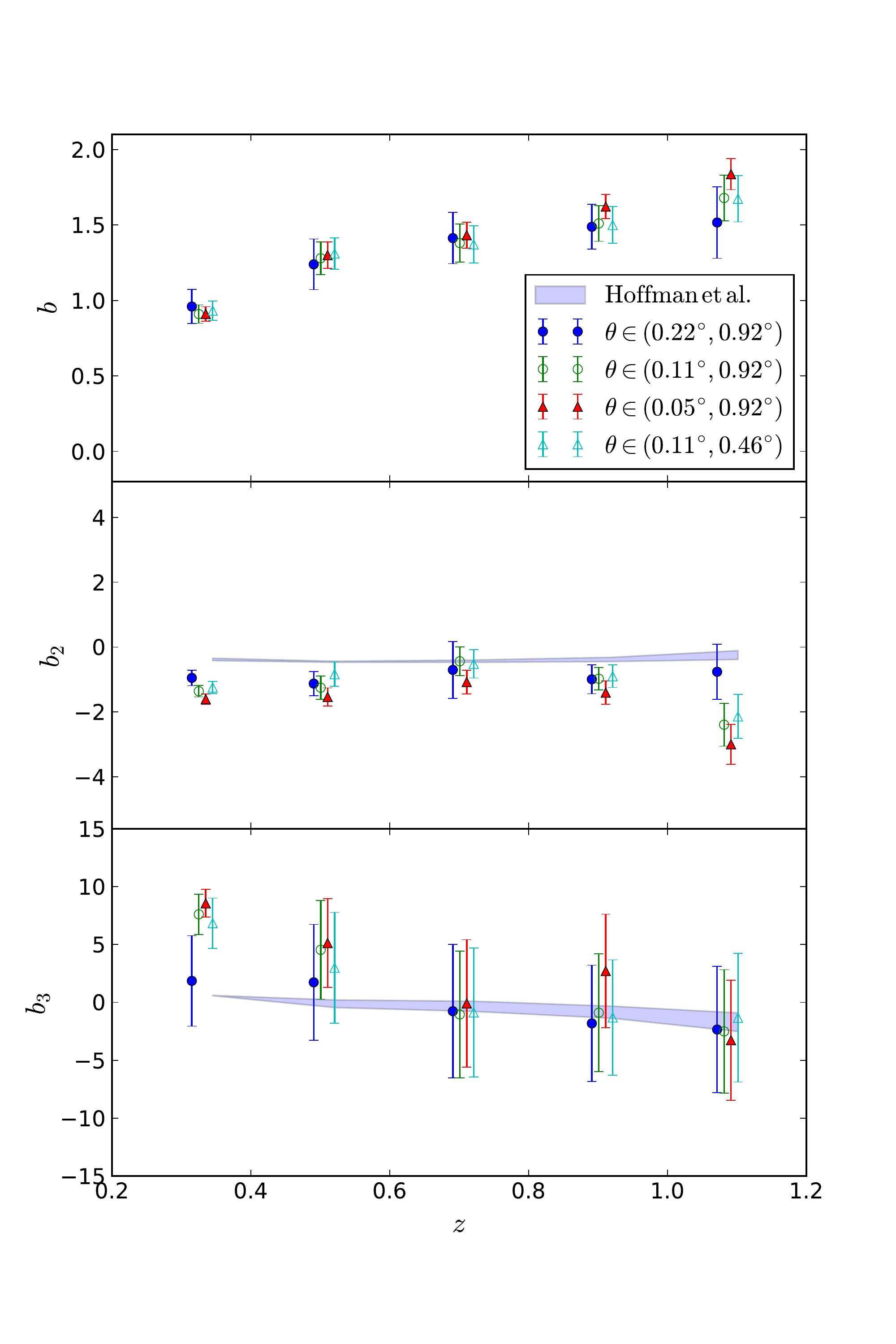}
\caption{Linear and non-linear bias results as a function of redshift for DES-SV data. Systematic uncertainties from Section \ref{sec:sys} are already included in these results, excluding the uncertainties associated to the modeling. The different marker shapes represent the best-fit results considering different ranges of aperture angle $\theta$. For the solid triangles we consider the range from $0.05^{\circ}$ to $0.92^{\circ}$, open circles symbolize our fiducial case with $0.11^{\circ} < \theta < 0.92^{\circ}$, in solid circles, we take out the smallest scale in our fiducial case and, in open triangles, we take out the largest scale. The shadowed region corresponds to the 3D predicted values using equations (\ref{eq:hoffman}). The top panel shows the projected linear bias $b$ as a function of redshift, the middle panel shows the best-fit results for the projected $b_{2}$ and $b_{3}$ is shown in the lower panel. The results for a given redshift $z$ have been separated in the figure for visualization purposes.}
\label{fig:bnl_bench_2}
\end{figure}

\subsection{Projected galaxy bias in DES - SV}
Repeating the procedure that we used for the MICE galaxy simulation, we analyze the DES - SV data and the MICE dark matter simulation, and compare their moments. In Figure \ref{fig:bnl_bench_2} we can see the results of simultaneously fitting for $b$, $b_{2}$ and $b_{3}$. The measurements in this figure include the systematic uncertainties are introduced in Section 6. The resulting $b$ is corrected by the ratio of $\sigma_{8}$ between MICE and our adopted fiducial cosmology using equation (\ref{eq:bias_corr}). The fit results can be seen in Figure \ref{fig:fits_data_gaussian_nl}. In this case, we detect a non-zero value for $b_{2}$. We check the probability of $b_{2}$ being zero by computing:
\begin{equation}
\chi^{2}_{z}=\sum_{i,j=1,N_{zbins}}\hat{b}_{2,i}\mathcal{C}_{2,ij}^{-1}(z)\hat{b}_{2,j}
\end{equation}
The sum runs for all the redshift bins. $\hat{b}_{2}$ is the weighted average of the fit results with the different fitting ranges and $\mathcal{C}_{2,ij}(z)$ is the covariance matrix for $b_{2}$. Taking into account the correlations between different redshift bins:
\begin{equation}
\mathcal{C}_{2,ij}(z)=\frac{N_{ij}N_{ji}}{N_{ii}N_{jj}}\Delta\hat{b}_{2,i}\Delta\hat{b}_{2,j}
\end{equation}
with $N_{ij}$ is the number of galaxies observed in the photo-z bin $i$ from the true-z bin $j$ and $\Delta\hat{b}_{2,i}$ is the weighted uncertainty in $\hat{b}_{2,i}$ for the photo-z bin $i$.
The value of $\chi^{2}_{z} = 64.75$ with 4 degrees of freedom, so the probability is essentially 0, making clear that the overall value of $b_{2}$ is non-zero for the local model. However, we lack the sensitivity necessary to detect a non-zero $b_{3}$. 

\begin{figure}
\centering
\includegraphics[scale=0.45,trim={0.8cm 0.5cm 0 1.5cm}, clip]{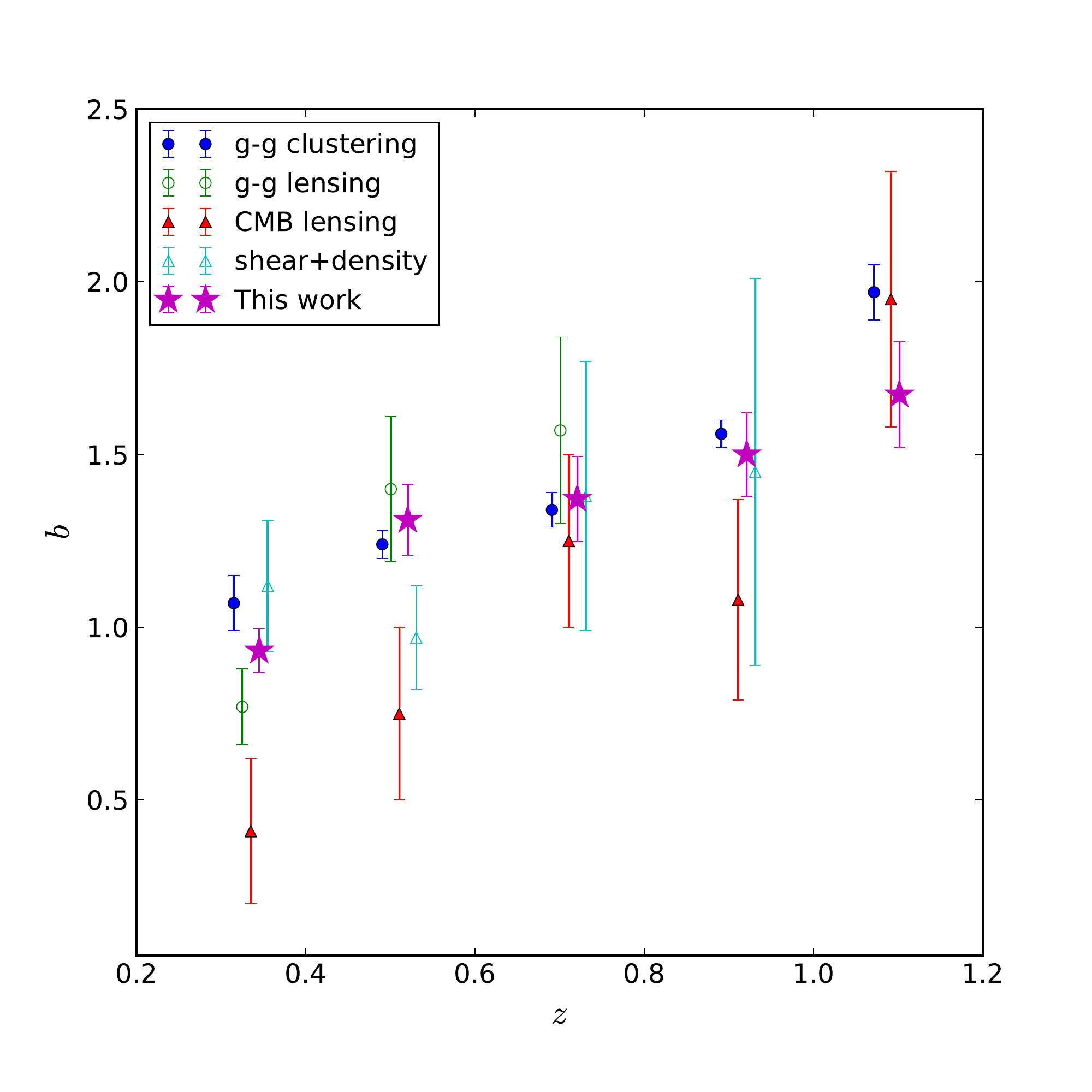}
\caption{Bias obtained from second order CiC, including systematic uncertainties from Section \ref{sec:sys}, compared with the 2-point correlation study \citep{2016MNRAS.455.4301C}, the CMB-galaxy cross-correlations study \citep{2016MNRAS.456.3213G}, galaxy-galaxy lensing \citep{2018MNRAS.473.1667P}, and the shear+density analysis~\citep{2016MNRAS.459.3203C}. The points for the same $z$ have been separated in the horizontal axis for visualization purposes.}
\label{fig:linear_bias_comparison}
\end{figure}

\noindent
We also check the measurement of linear bias obtained in this work and compare it with previous measurements on the same dataset Figure~\ref{fig:linear_bias_comparison}. The measurements are generally in good agreement with each other showing the robustness of the method.

\noindent
Future DES data will have a considerably larger area and, as previous MICE measurements show, these measurements will improve. Here we also use the skewness and the kurtosis of dark matter from the MICE dark matter simulation, as those quantities hardly depend on the cosmology~\citep{Bouchet:1992uh}. We also find that our results are similar to those in \citet{Ross:2006zr}. We do not expect them to be equal as the samples are different and the bias depends strongly on the population sample.

%--------------------------------------------------------------------------------------
%--------------------------------------------------------------------------------------
\section{Systematic Errors}\label{sec:sys}

\begin{figure*}
\centering
\includegraphics[width=0.9\textwidth]{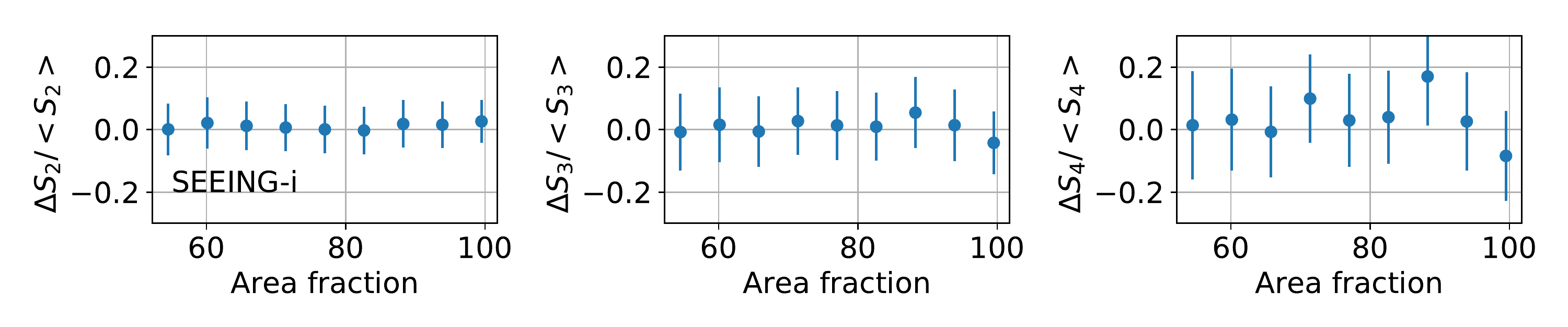}
\includegraphics[width=0.9\textwidth]{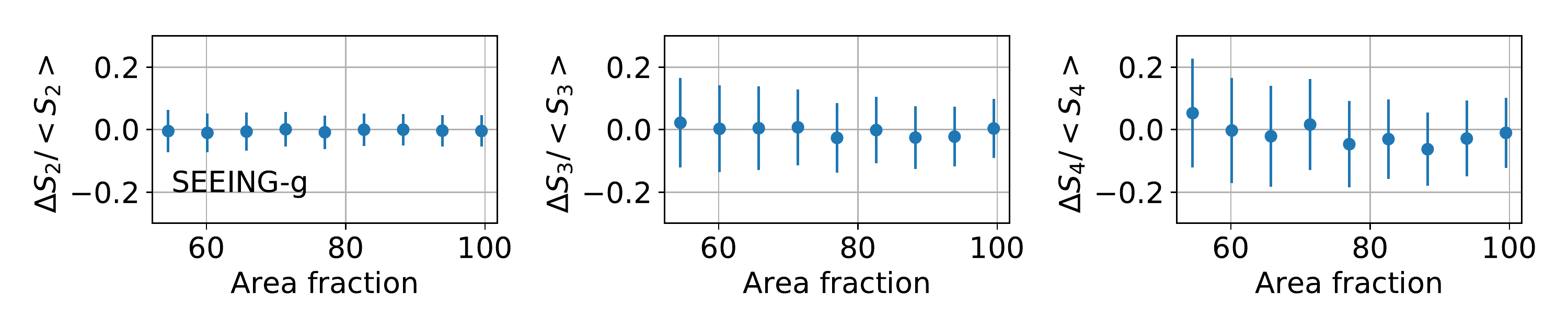}
\includegraphics[width=0.9\textwidth]{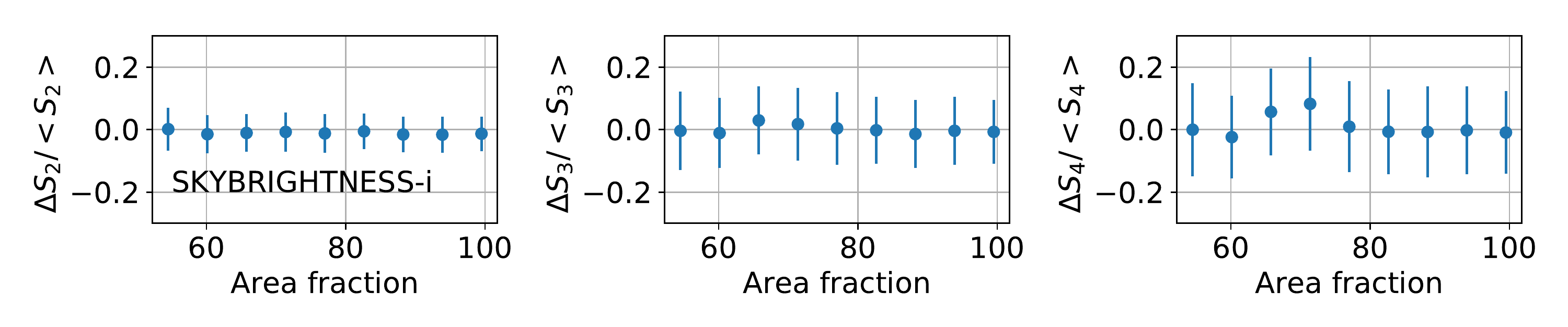}
\includegraphics[width=0.9\textwidth]{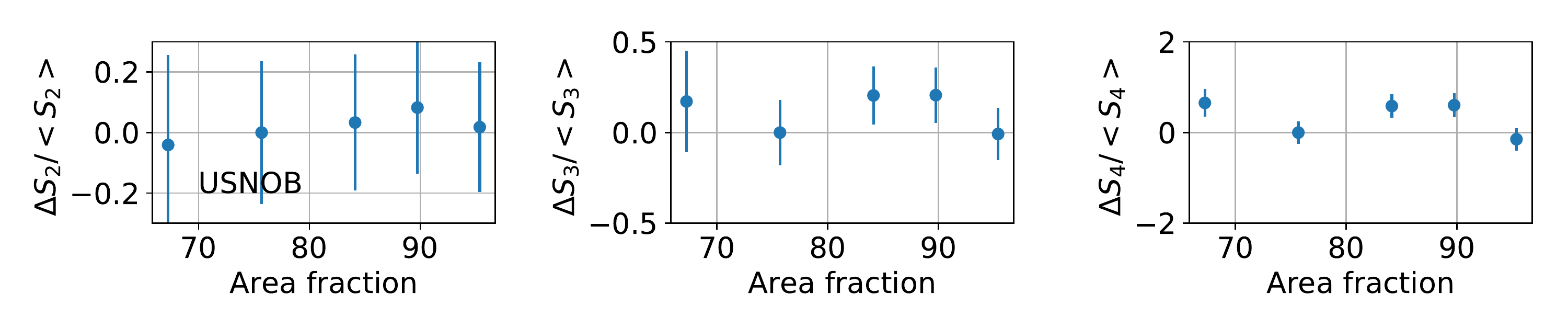}
\caption{Dependence of the moments $S_{i}$ with the variation in the value of potential systematic effects. We show an example for $N_{side}=2048$ in the redshift bin $0.2 < z < 0.4$ using TPZ. The left column shows the behavior for $S_{2}$, the middle column shows $S_{3}$, and the last column shows the results for $S_{4}$. The first row corresponds to the results for the seeing in i-band, the second row shows the results for seeing in g-band, the third shows the sky-brightness in i-band. Finally the last row shows the evolution of the moments with the variation in the number of stars per pixel.}
\label{fig:sys_example}
\end{figure*}

In this section, we explore the effects that several potential sources of systematic uncertainty have on our moment measurements. Since our main observable is related to the number of galaxy-counts in a given redshift interval, we are interested in observational effects that can affect this number. The main potential sources of systematic uncertainties are changes in airmass, seeing, sky brightness, star-galaxy separation, galactic extinction, and the possible errors in the determination of the photometric redshift. In order to evaluate their effects, we use the maps introduced in \citet{2016ApJS..226...24L}. To account for the stellar abundance in our field we proceed as in \citet{2016MNRAS.455.4301C} and use the USNO-B1 catalog \citep{2003AJ....125..984M}. We also use the SFD dust maps ~\citep{1998ApJ...500..525S}.
\noindent
What follows is a detailed step-by-step guide to our systematic analysis: we select one of the aforementioned maps and locate the pixels where the value of the systematic is below the percentile level $t$. We compute the moments of the density contrast distribution in these pixels and their respective errors using bootstrap. We change the threshold to $t+5$, repeat the process, and evaluate the difference between the moments calculated using this threshold divided by the moments in the original footprint $\Delta S_{i}(t)$/$\langle S_{i} \rangle$. An example of the results of this procedure can be found in Figure \ref{fig:sys_example}. Note that the plot showing the variation of the moments with USNOB shows less points in the horizontal axis. Due to the discrete nature of the map of stellar counts, the 50th and 60th percentiles of the $\delta$ distribution of the stellar counts are the same, in order to avoid these problems, we make less bins in this case.

\noindent
We consider that a systematic effect is present if the average of $\Delta S_{i}(t)$/$\langle S_{i} \rangle$ is different from zero at a $2\sigma$ confidence level or above for the different values of $t$ from the 50th tile to the 100th tile. Then, we assign a systematic uncertainty equal to the value of this average. To be conservative, we consider these effects as independent, so we add them in quadrature. We summarize the main systematic effects observed in each redshift bin of our sample:
\begin{itemize}
\item Bin $0.2 < z < 0.4$:
\begin{itemize}
\item Seeing in i-band: we assign a 3\% systematic uncertainty in $S_{4}$.
\item Seeing in z-band: we assign a 2.5\% systematic uncertainty in $S_{4}$.
\item Sky-brightness r-band: we assign a 1\% systematic uncertainty in $S_{4}$.
\item Sky-brightness i-band: we assign a 1\% systematic uncertainty in $S_{4}$.
\item Airmass in g-band: we assign a 1\% uncertainty in $S_{4}$.
\item Airmass in r-band: we assign a 1\% uncertainty in $S_{4}$.
\item Airmass in i-band: we assign a 1\% uncertainty in $S_{4}$.
\item USNO-B stars: We assign a 4\% uncertainty to $S_{2}$, 7\% uncertainty to $S_{3}$, and 9\% to $S_{4}$.
\end{itemize}
\item Bin $0.4 < z < 0.6$:
\begin{itemize}
\item Seeing in z-band: We assign a 1.5\% uncertainty to $S_{4}$.
\item USNO-B stars: We assign a 4\% uncertainty to $S_{2}$, 3\% uncertainty to $S_{3}$, and 4\% to $S_{4}$.
\end{itemize}
\item Bin $0.6 < z < 0.8$:
\begin{itemize}
\item Seeing in g-band: We assign a 2\% to $S_{4}$.
\item Seeing in r-band: We assign a 2\% to $S_{4}$.
\item Sky-brightness i-band: We assign a 1.5\% uncertainty to $S_{3}$, and 3\% systematic uncertainty to $S_{4}$.
\item Airmass in g-band: We assign a 2.5\% uncertainty to $S_{4}$.
\item Airmass in r-band: We assign a 2\% uncertainty to $S_{4}$.
\item Airmass in z-band: We assign a 1.5\% uncertainty to $S_{3}$, and 3\% uncertainty to $S_{4}$.
\item USNO-B stars: We assign a 3\% uncertainty to $S_{3}$, and 5\% uncertainty to $S_{4}$.
\end{itemize}
\item Bin $0.8 < z < 1.0$:
\begin{itemize}
\item Seeing in g-band: We assign a 2\% uncertainty to $S_{4}$.
\item Sky-brightness in i-band: We assign a 2\% uncertainty to $S_{3}$, and a 3.5\% uncertainty to $S_{4}$.
\item Airmass in g-band: We assign a 2\% uncertainty to $S_{4}$.
\item Airmass in r-band: We assign a 3\% uncertainty to $S_{4}$.
\item USNO-B stars: We assign a 3\% uncertainty to $S_{4}$.
\end{itemize}
\item Bin $1.0 < z < 1.2$:
\begin{itemize}
\item The measurement of $S_{4}$ in this bin is dominated by systematics.
\item Sky-brightness i-band: We assign 2\% to $S_{3}$.
\item Sky-brightness z-band: We assign 3\% to $S_{3}$.
\item USNO-B stars: We assign a 4.5\% uncertainty to $S_{3}$.
\end{itemize}
\end{itemize}
\noindent
The estimated systematic errors for the bias are propagated from the estimation of the systematics in $S_2$, $S_3$, and $S_4$. Their behavior is compatible with the systematics found in \citet{2016MNRAS.455.4301C}. We use the same data masking, excluding regions with large systematic values to recover $w(\theta)$. The linear bias is more robust using CiC since the variance, $S_{2}$, is less affected by the small scale power induced by the systematics given that these scales are smoothed out. On the other hand, the non-linear bias is more sensitive to the presence of systematics because they can induce asymmetries in the density contrast distribution. 

\subsection{Photometric redshift}

\begin{figure}
\includegraphics[scale=0.45]{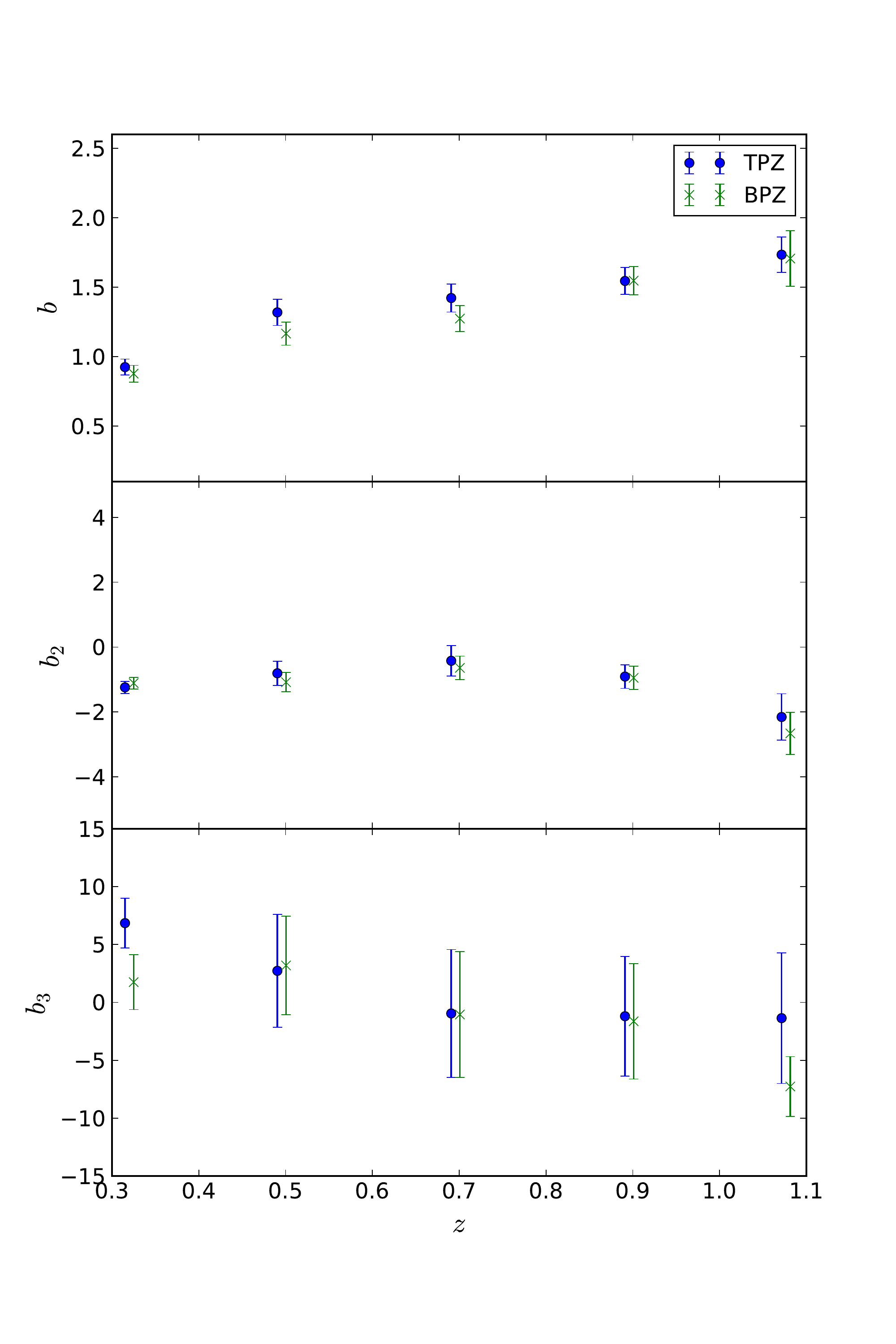}
\caption{Bias obtained in the SV data from second order CiC for TPZ (solid blue circles) and BPZ (green crosses). The results for a given redshift $z$ have been separated in the figure for visualization purposes.}
\label{fig:biasbenchbpz}
\end{figure}

Photometric redshift is one of the main potential sources for systematic effects in photometric surveys like DES. We have repeated the analysis in DES-SV data for a second estimate of the photometric redshift using BPZ ~\citep{Benitez:1998br}. In Figure \ref{fig:biasbenchbpz} we compare the results for the two photometric redshift codes and we see that they are in good agreement. The linear bias seems to be the most affected by the choice of a photometric redshift estimator but the results do not show any potential systematic biases. For the non-linear bias we get remarkably consistent results, showing the robustness of this method.

%--------------------------------------------------------------------------------------
%--------------------------------------------------------------------------------------
\subsection{Biasing models}
\begin{figure}
\includegraphics[scale=0.45]{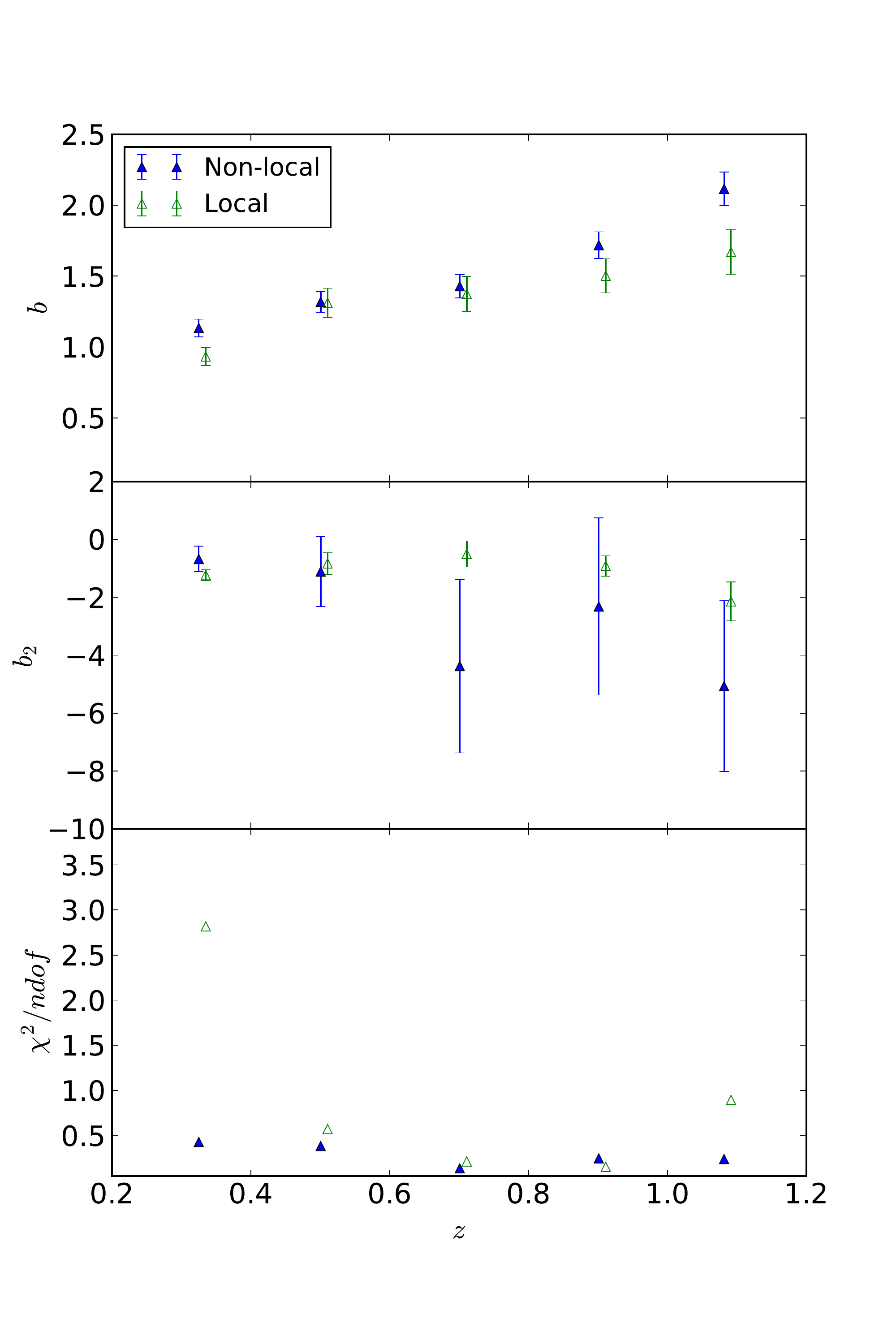}
\caption{(Top) Comparison between the linear bias results obtained with CiC for SV using different biasing models: non-local (solid triangles), and local (open triangles) using the TPZ sample. (Middle) Comparison between $b_{2}$ results for the same models as above. (Bottom) total reduced chi square for each of the models.}
\label{fig:model_comp_sv}
\end{figure}

Apart from the terms that we considered in our model, \citet{Bel:2015jla} found that non-local bias terms are responsible for the overestimation of the linear bias from the three-point correlation in \citet{Pollack:2013alj,Hoffmann:2015mma,2011MNRAS.415..383M} but that they should not significantly affect second-order statistics. As we mentioned previously in Section 5, we do not expect these terms to have a significant impact on our estimations because we analyze projected quantities over considerable volumes (note that we integrate in the cell and in the redshift slice). Having said that, we test the local and non-local models and find the results depicted in Figure \ref{fig:model_comp_sv}. We can see, as in the case of the simulation, that both models are consistent within errors. This means that choosing the local model does not introduce any systematic uncertainties in our linear bias measurements. However, it affects the $b_{2}$ measurements and their uncertainty since the new parameters introduced with these more complicated models are correlated with them. We check the probability of $b_{2}$ being zero for the different models and obtain the results in table \ref{tab:bias_models}. We find $b_{2}$ to be different from zero at a 3-$\sigma$ level in the worst case (non-local).
\begin{table}
\centering
\begin{tabular}{|c|c|c|c|}
\hline
Bias model & $\chi^{2}$ & $p$-value & ndof\\
\hline
Local & 64.75 & $3\times 10^{-13}$ & 4\\
Non-local & 12.63 & 0.013 & 4\\
\hline
\end{tabular}
\caption{Comparison on the null hypothesis for $b_{2}$ in DES-SV data for the different bias models considered in this work.}
\label{tab:bias_models}
\end{table}
We also can see that in the first bin, none of the models fit the data well, which is not surprising, given that the range of (comoving) scales is very small ($\sim 1 - 20$ Mpc $h^{-1}$) and non-linear clustering dominates.

Finally, we are not considering stochastic models and we are assuming a Poisson shot-noise. This means that our measured $b_{2}$ could be entangled with stochasticity~\citep{1998ApJ...504..601P,2013PhRvD..87l3523S}. We leave the study of stochasticity to future works. 
\subsection{Value of $\sigma_{8}$}
\label{ssec_sys_s8}
As mentioned in previous sections, our bias estimation depends linearly on the value of $\sigma_{8}$. Thus, if the actual value of $\sigma_{8}$ is different from our assumed fiducial value, our results will be biased, and we have to correct for the difference using equation \ref{eq:bias_corr}. This is why, we introduce a systematic uncertainty of $1.4\%$ (the uncertainty level in $\sigma_{8}$ from \citet{Ade:2013zuv}) that we add in quadrature to the statistical errors in the final estimation of the bias.

\section{Conclusions}\label{sec:conclusions} 

CiC is a simple but effective method to obtain the linear and non-linear bias. A good measurement of the galaxy bias is essential to maximize the performance of photometric redshift surveys because it can introduce a systematic effect on the determination of cosmological parameters. The galaxy bias is highly degenerate with other cosmological parameters and an independent method to determine it can break these degeneracies and improve the overall sensitivity to the underlying cosmology. In this paper we have developed a method to extract the bias from CiC. We use the MICE simulation to test our method and then perform measurements on the public Science Verification data from the Dark Energy Survey. The strength of this method is that it is based on a simple observable, the galaxy number counts, and is not demanding computationally. 

We check that our linear bias measurement from CiC agrees with the real bias in the MICE simulation. Figure \ref{fig:linear_bias_mice} shows an agreement between our measurement and the one obtained using the angular two-point correlation function. We then obtain the linear bias in the SV data and find that it is in agreement with previous bias measurements from other DES analyses. In Figure \ref{fig:linear_bias_comparison}, we see that the CiC values are compatible with the two-point correlation study~\citep{2016MNRAS.455.4301C}, the CMB-galaxy cross-correlations study~\citep{2016MNRAS.456.3213G}, and the galaxy-galaxy lensing~\citep{2018MNRAS.473.1667P}, and we demonstrate that these results are robust to the addition of new parameters in the biasing model, such as the non-local bias. Finally, we compute the non-linear bias parameters up to third order. We detect a significant non-zero $b_{2}$ component. It appears that the 2D and 3D predictions of the non-linear bias are in better agreement at larger scales, as expected. However, given the uncertainties associated with these quantities, it is difficult to draw any conclusions from $b_{3}$ despite its compatibility with the expected 3D prediction. When more data is available, we plan to check if we can improve our constraints on $b_{3}$ and whether the agreement with the 3D prediction improves as well. The systematic errors are in general lower than the statistical errors, in agreement with the systematic study done by~\citep{2016MNRAS.455.4301C}.
 
%--------------------------------------------------------------------------------------
%--------------------------------------------------------------------------------------
\appendix

\section{Different pixel shapes}\label{sec:appendix_b}
\begin{figure}
\includegraphics[scale=0.45]{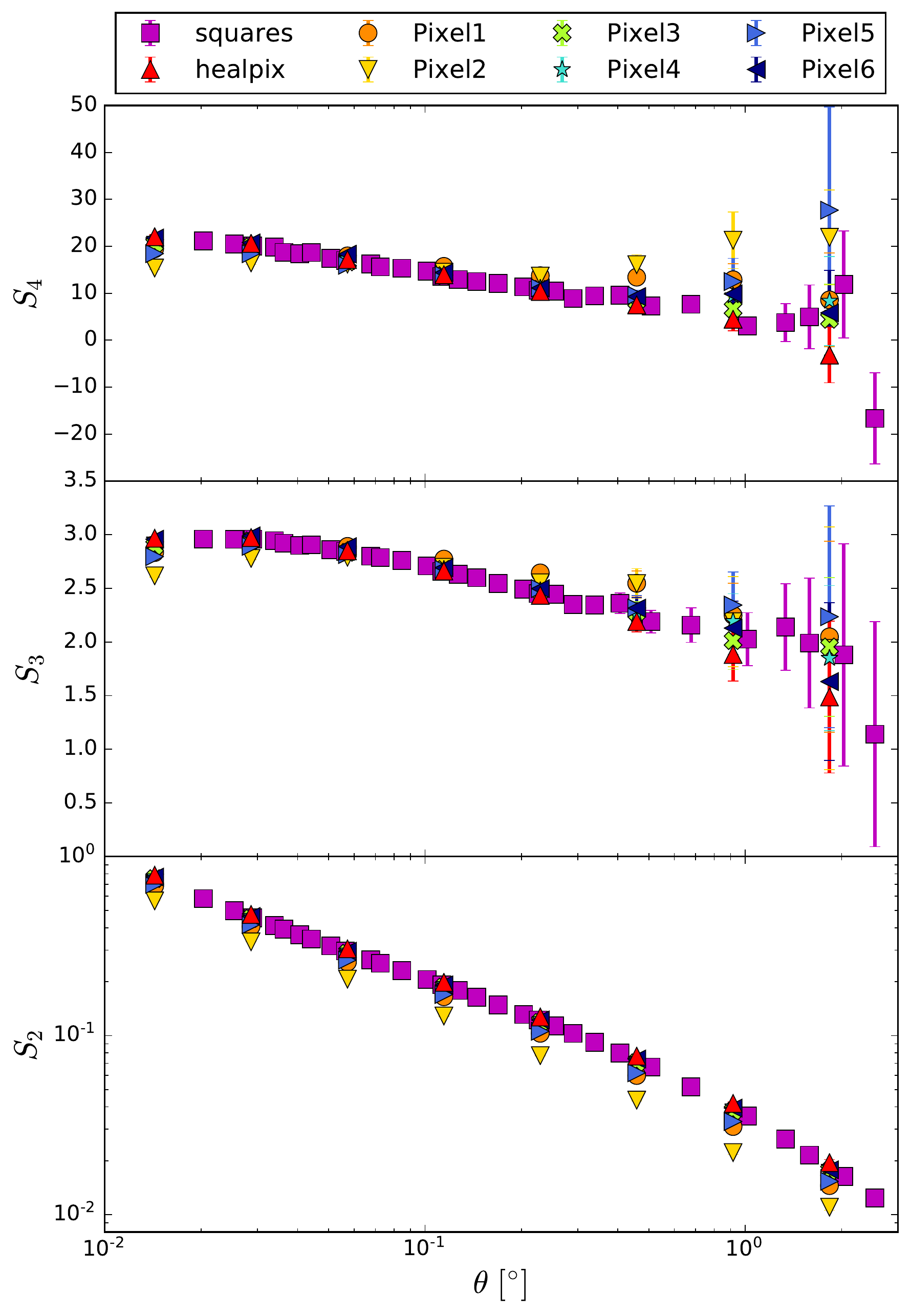}
\caption{Moments of the density contrast distribution as a function of the cell scale using data from MICE in the redshift slice $0.95<z<1.05$ for different pixel shapes. Pixels number 3 ($n_{ra}=3N_{side},\ n_{ct}=4N_{side}$), 4 ($n_{ra}=4N_{side},\ n_{ct}=3N_{side}$) and 6 ($n_{ra}=6N_{side},\ n_{ct}=2N_{side}$) are close to being squares, but pixels number 1 ($n_{ra}=12N_{side},\ n_{ct}=1N_{side}$), 2 ($n_{ra}=1N_{side},\ n_{ct}=12N_{side}$) and 5 ($n_{ra}=2N_{side},\ n_{ct}=6N_{side}$) are far from being regular polygons.}
\label{fig:shapes}
\end{figure}

We check with the MICE simulation in a thin redshift bin ($0.95<z<1.05$) that as long as we have regular polygon pixels the difference in the moments of the density contrast is negligible. In Figure \ref{fig:shapes} we see that the difference is negligible for the more symmetrical pixels and higher for less symmetrical ones. The angular aperture, $\theta$, is estimated as the square root of the pixel area. We compare rectangular pixels with \texttt{HEALpix} pixels. We divide the sphere into rectangular pixels taking $n_{ra}$ parts in right ascension, and $n_{ct}$ parts in $\sin \mathrm{dec}$ where the number of pixels is npix=$n_{ra}\cdot n_{ct}=12\cdot N_{side}\cdot N_{side}$. We have taken six different pixel shapes numbered from 1 to 6. Pixels number 3 ($n_{ra}=3N_{side},\ n_{ct}=4N_{side}$), 4 ($n_{ra}=4N_{side},\ n_{ct}=3N_{side}$) and 6 ($n_{ra}=6N_{side},\ n_{ct}=2N_{side}$) are close to being squares, but pixels number 1 ($n_{ra}=12N_{side},\ n_{ct}=1N_{side}$), 2 ($n_{ra}=1N_{side},\ n_{ct}=12N_{side}$) and 5 ($n_{ra}=2N_{side},\ n_{ct}=6N_{side}$) are far from being regular polygons. When we compare square and \texttt{HEALpix} pixels we see that the measured moments are in perfect agreement.
%Appendix C
\section{Boundary Effects}
\label{sec:appendix_c}
\begin{figure}
\includegraphics[scale=0.45]{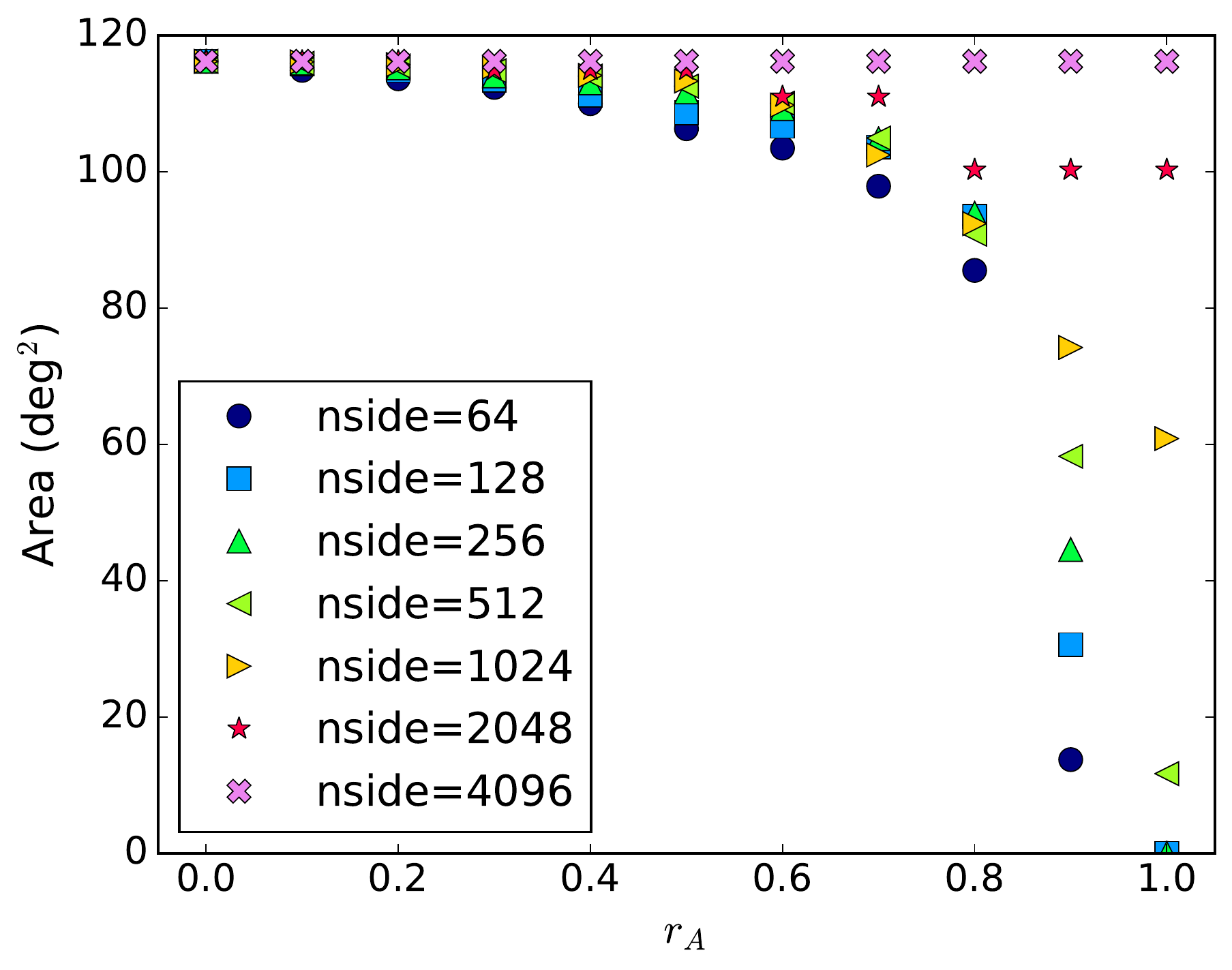}
\caption{Area covered by different \texttt{HEALpix} pixelation resolutions as a function of the minimum fraction of pixel coverage of said resolution with respect to the $N_{side}=4096$ footprint (larger pixels from lower $N_{side}$ will be partially filled at times). This test is done using the MICE simulation considering the same footprint as the SV dataset.}
\label{fig:area_cut}
\end{figure}

\begin{figure}
\centering
\includegraphics[scale=0.35]{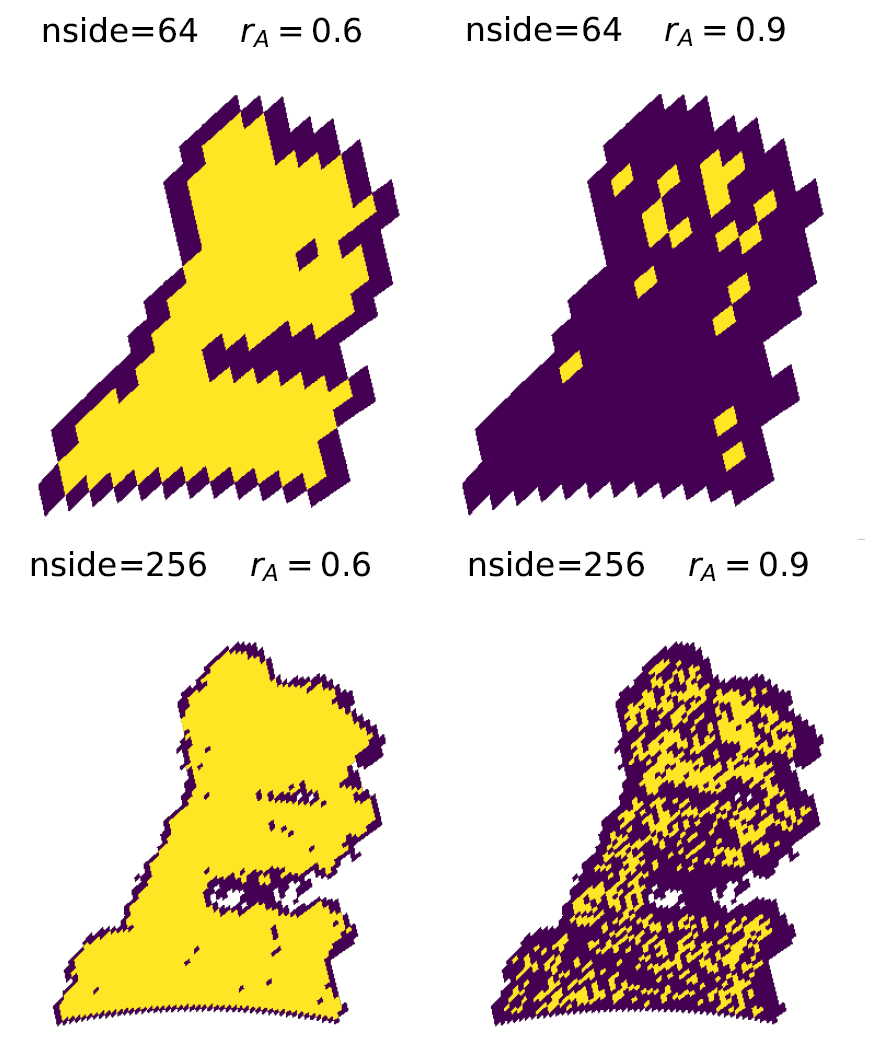}
\caption{DES SV mask for different $N_{side}$ (64, 256) and different area cuts $r_A=0.6,\ 0.9$. The pixels that we discard are blue and the ones that we keep are red. The bigger the pixel, the larger the amount of data we lose.}
\label{fig:areamask}
\end{figure}

\begin{figure}
\includegraphics[scale=0.45]{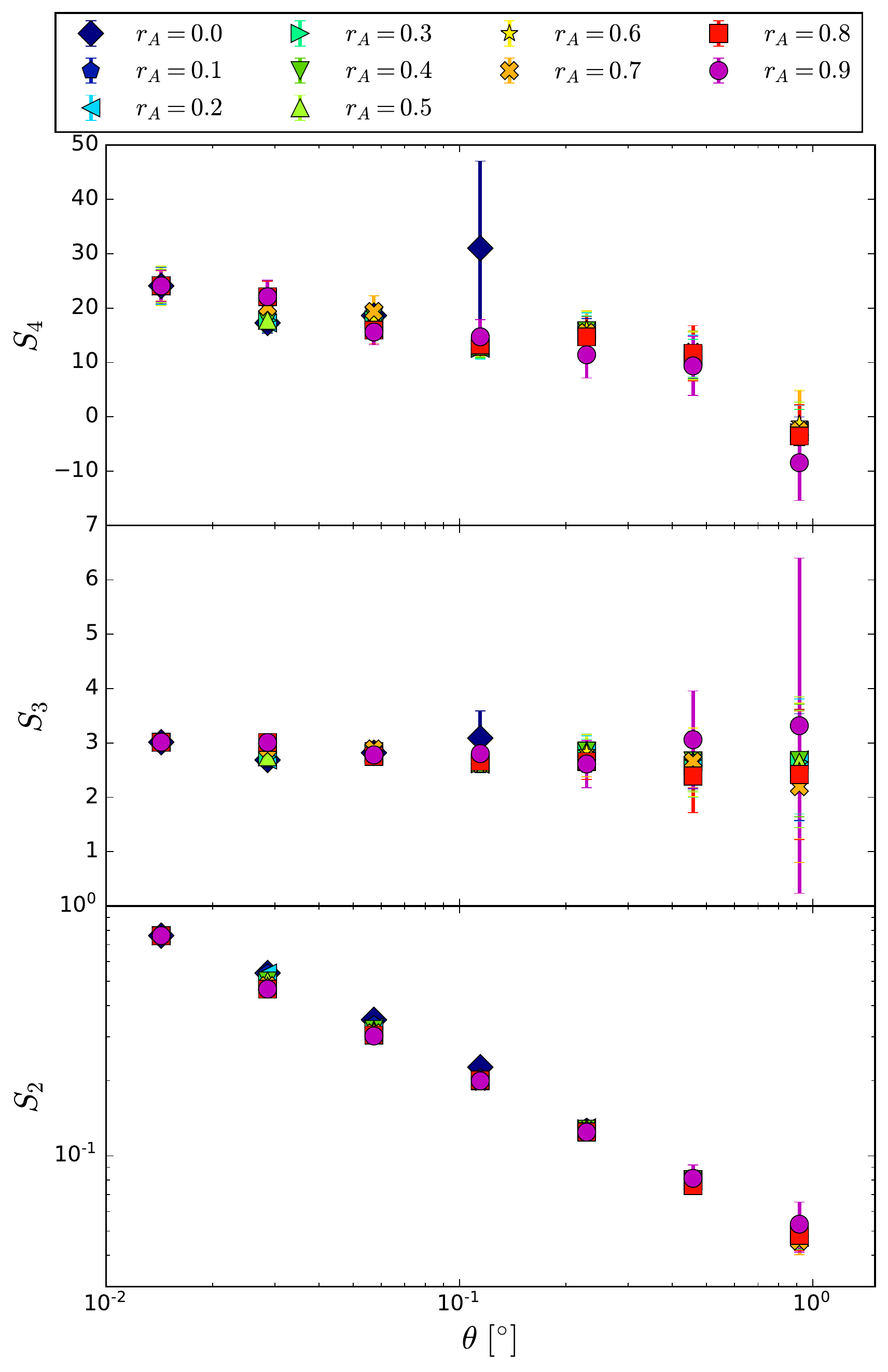}
\caption{Moments of the density contrast distribution obtained from MICE ($0.95<z<1.05$) considering the same footprint as the SV data for different values of the fraction of the pixel inside the mask, $r_{A}$. The results for a given scale $\theta$ have been separated in the figure for visualization purposes.}
\label{fig:areamom}
\end{figure}

To deal with the boundary effects of an irregularly shaped area, we use the mask and degrade its resolution to match each of the pixel scales being used. However, degrading the mask (or increasing the scale) results in an increasing number of partially filled pixels. Only a fraction $r_{A}=A_{\rm{filled}}/A_{\rm{pixel}}$ remain completely inside the footprint. This means that, if we assign the same scale to all the pixels of a given $N_{side}$ value, some pixels will be effectively mapping a different scale. To solve this problem we can either require a minimum fraction of the pixel to be full, $r_{A} \geq X$, or we can compute the fraction of full pixels and perform CiC for that scale. We prefer to use the former because we consider that the scales where we perform the study appropriately map the variations of the density field in which we are interested. This approach also helps to avoid certain boundary effects. For small pixel sizes (similar to the size in the mask), given the large number of pixels, we can safely choose $r_{A}=1$. For bigger pixels we try to find a compromise between the amount of area that we lose and the boundary effects. In Figures \ref{fig:area_cut} and \ref{fig:areamask} we show the area loss using data from MICE in the redshift bin $0.95<z<1.05$ with the SV mask for different thresholds in $r_{A}$ and in Figure \ref{fig:areamom} the change in the moments for these different area cuts. We see that if we choose pixels that are completely contained inside the mask ($r_{A} = 1.0$), we lose a lot of area for smaller values of $N_{side}$, however, very little area is lost for large values of $N_{side}$. It can be seen that results are consistent for the different threshold values for $r_{A}$. We also see that if we take all the pixels ($r_A\geq0$), the difference in the moments is considerable in some cases, and we cannot take just all the pixels inside the mask ($r_A=1$) because we run out of them for large scales. We set a threshold $r_{A} \geq 0.9$ to ensure that the pixels are almost completely embedded in the footprint. This prevents us from mixing scales even for the largest pixel sizes. This can be noted in Figure \ref{fig:area_cut} where a large drop in area occurs between $r_{A}=0.8$ and $r_{A}=0.9$ for $N_{side} \leq 1024$, setting this threshold naturally. For most scales this threshold does not change the errors. By choosing $r_A\geq0.9$ the effective cell sizes are well determined and the errors are reasonably small.

\section{Simultaneous fits results}
\label{app:fits}
In this section we show the fitting results for the simultaneous fits in MICE. In Figures D1 and D2, the red line corresponds to the mean value of the samples and the grey lines are the different models evaluated by the MCMC.
\begin{figure*}
\centering
\includegraphics[scale=0.35]{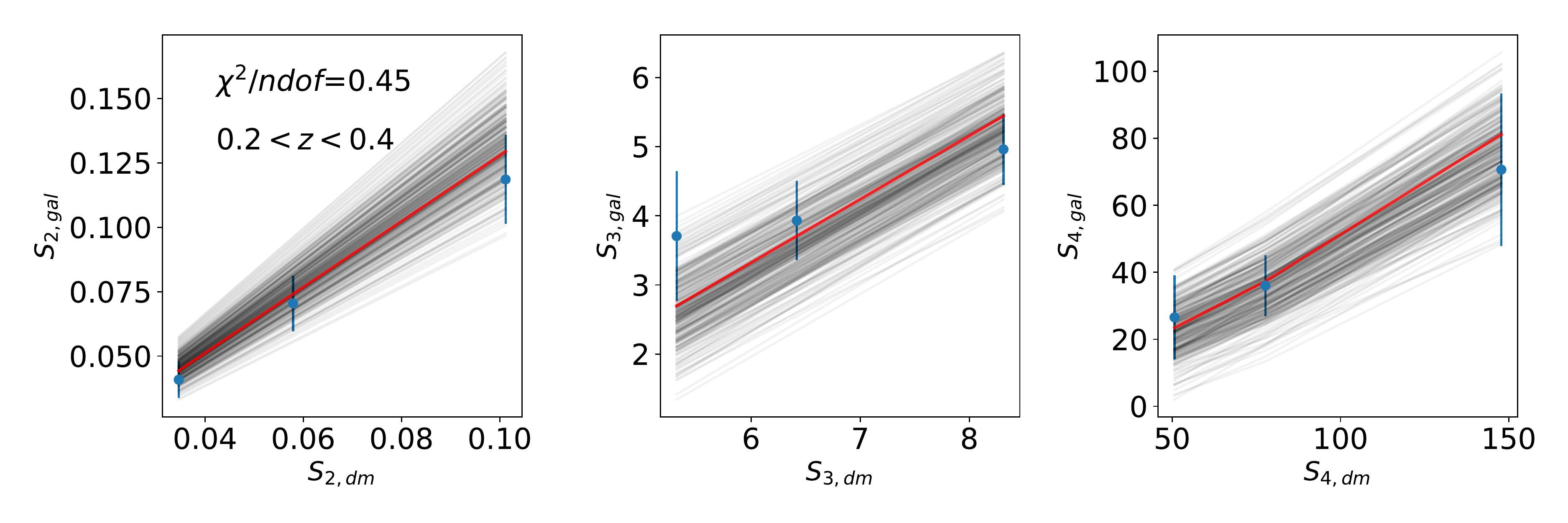}
\includegraphics[scale=0.35]{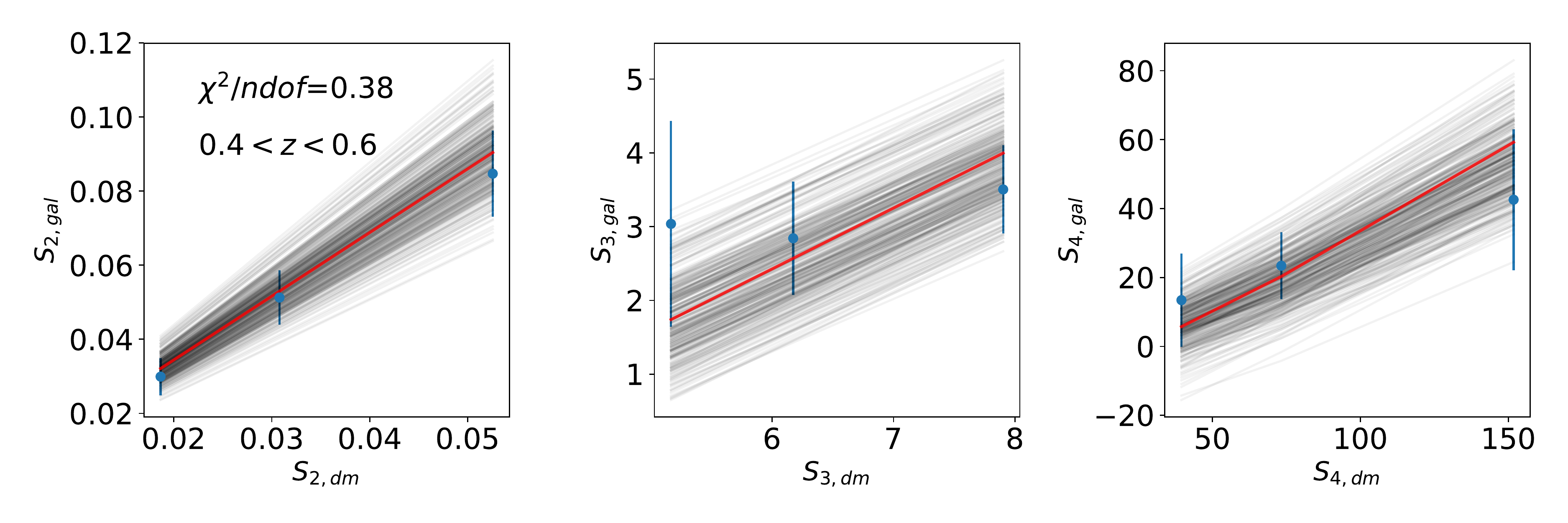}
\includegraphics[scale=0.35]{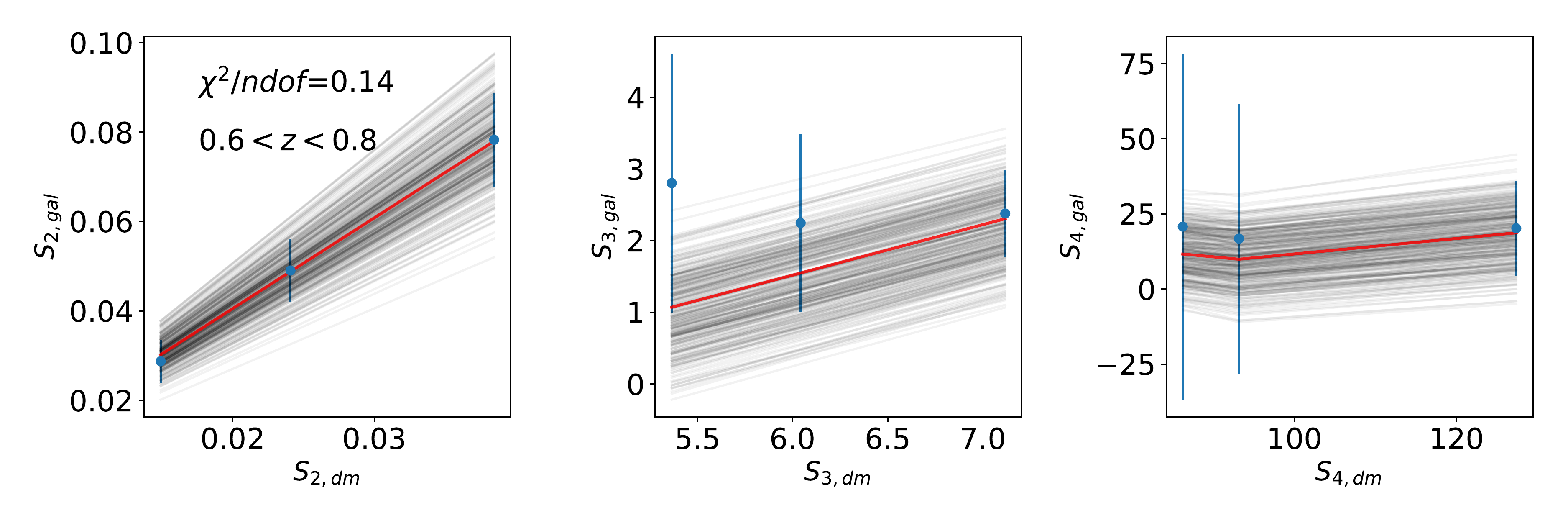}
\includegraphics[scale=0.35]{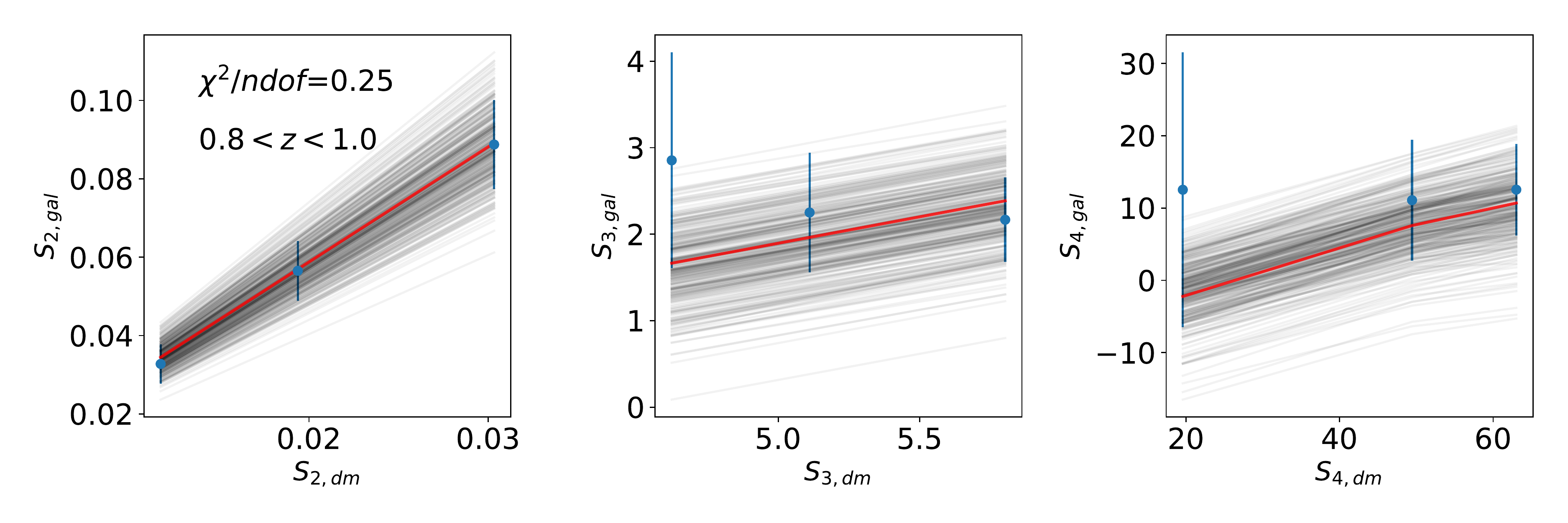}
\includegraphics[scale=0.35]{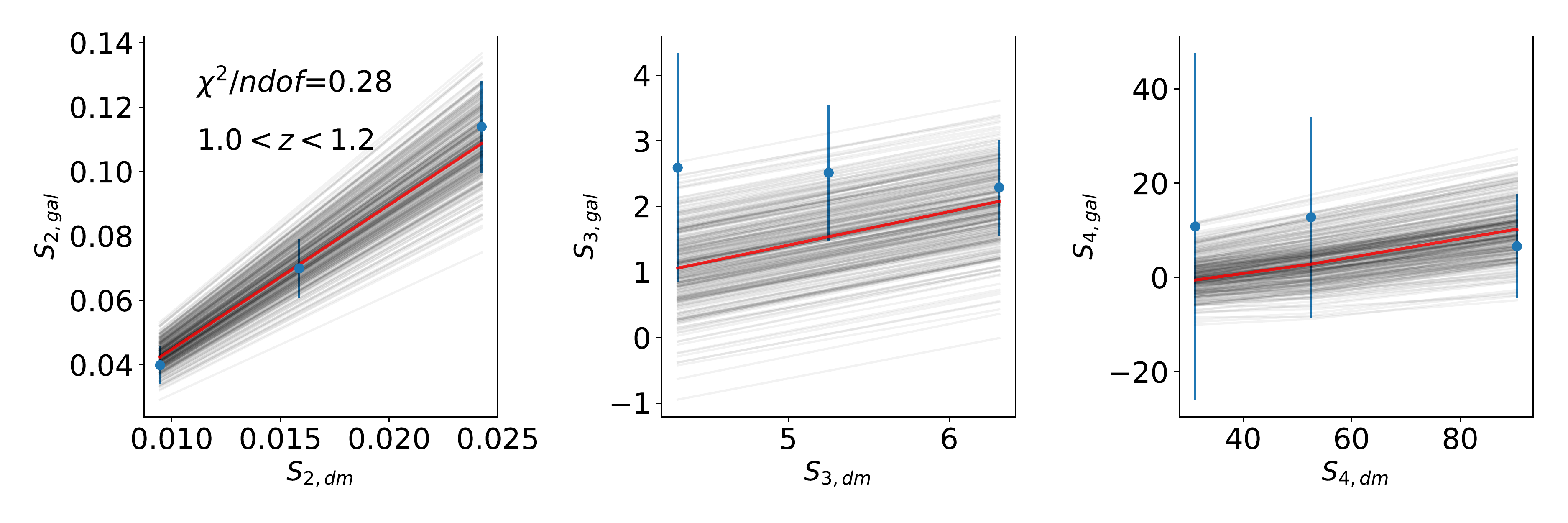}
\caption{Fit results for the non-linear bias simultaneous fits method using MICE with Gaussian photo-z. The points are the measured moments and the error bars are calculated by adding in quadrature the uncertainties from the moments in the dark matter and the galaxies. The thick dark line is the best-fit curve corresponding to the mean of the posterior distribution. The thin gray lines are the different models evaluated by the MCMC. The top row corresponds to the first redshift bin $(0.2 < z < 0.4)$, the second row corresponds to the second redshift bin, and so on.}
\label{fig:fits_mice_gaussian_nl}
\end{figure*}
%\begin{figure*}
%\centering
%\includegraphics[scale=0.45,angle=90,trim={6cm 0.3cm 6cm 0.3cm},clip]{mcmc/fit_bin_3_mice.pdf}
%\includegraphics[scale=0.45,angle=90,trim={6cm 0.3cm 6cm 0.3cm},clip]{mcmc/fit_bin_4_mice.pdf}
%\caption{See caption in Figure \ref{fig:fits_mice_gaussian_nl}.}
%\label{fig:fits_mice_gaussian_nl_2}
%\end{figure*}
\begin{figure*}
\centering
\includegraphics[scale=0.35]{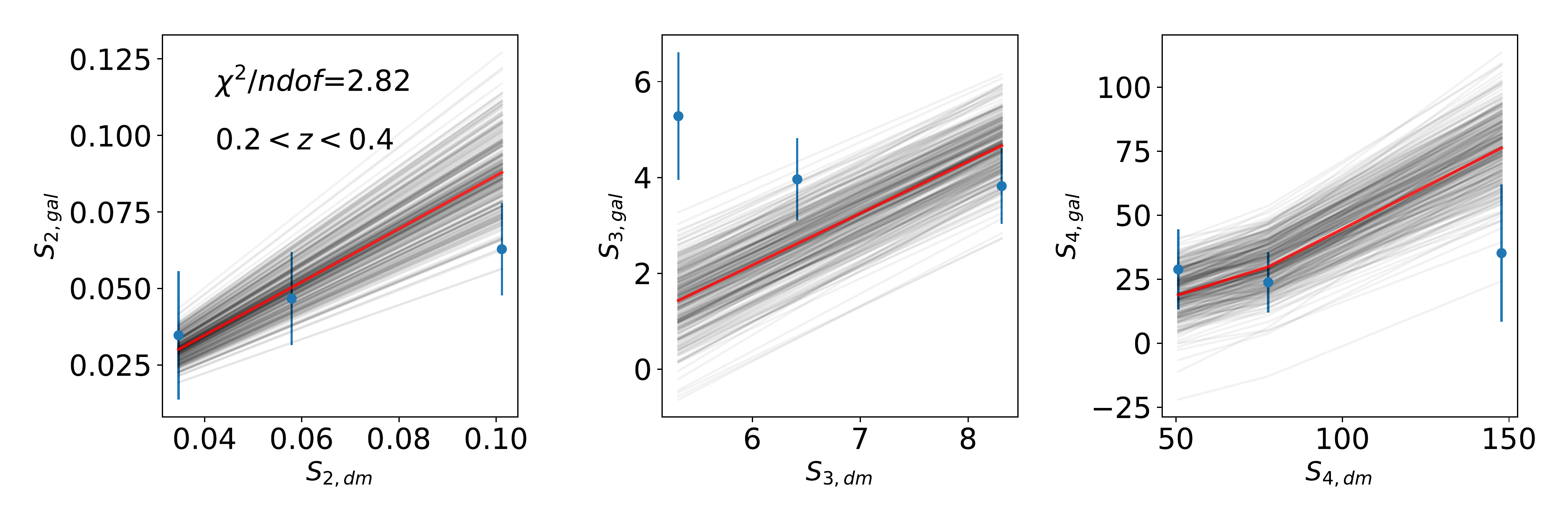}
\includegraphics[scale=0.35]{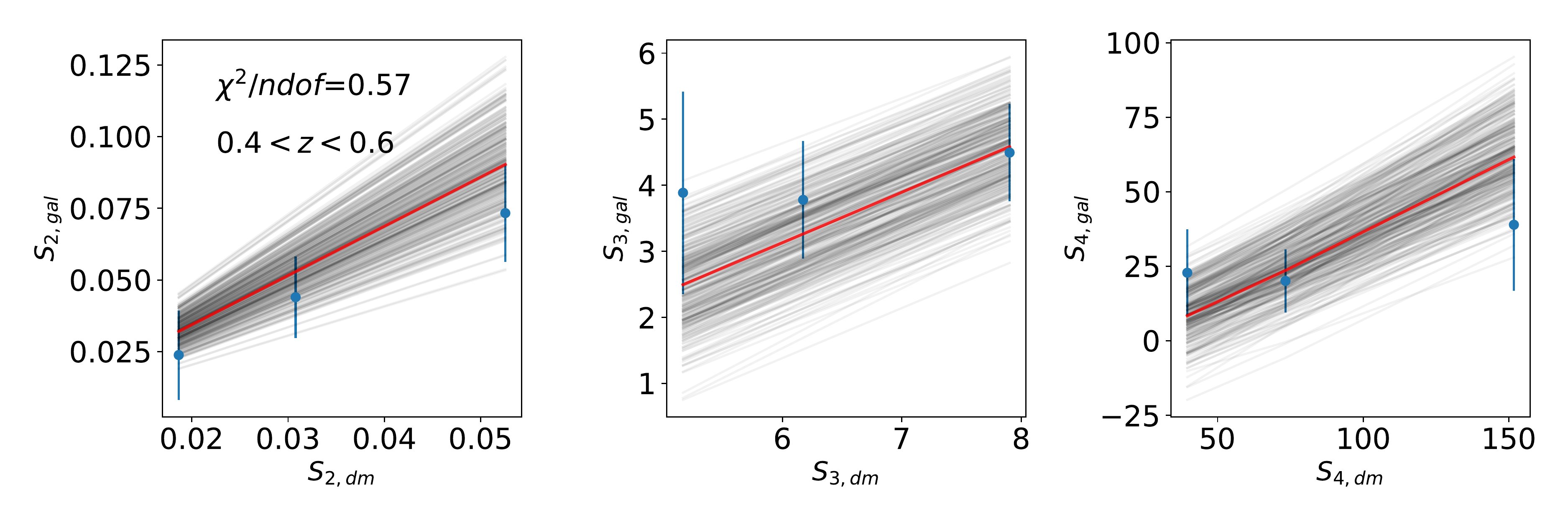}
\includegraphics[scale=0.35]{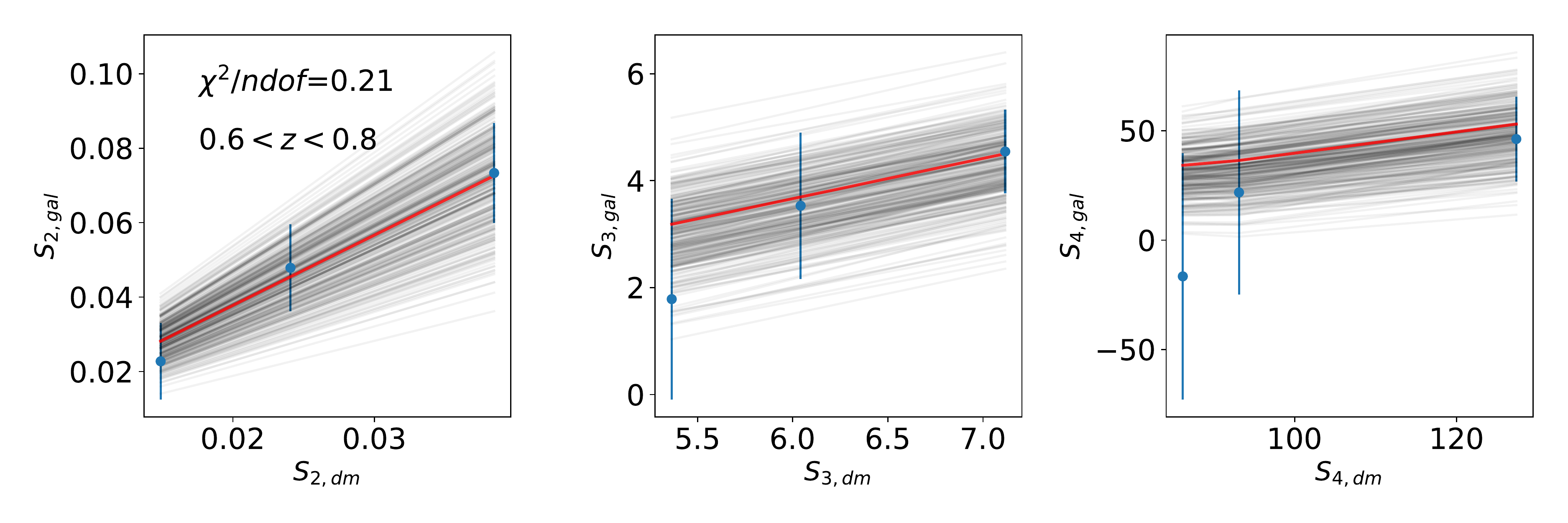}
\includegraphics[scale=0.35]{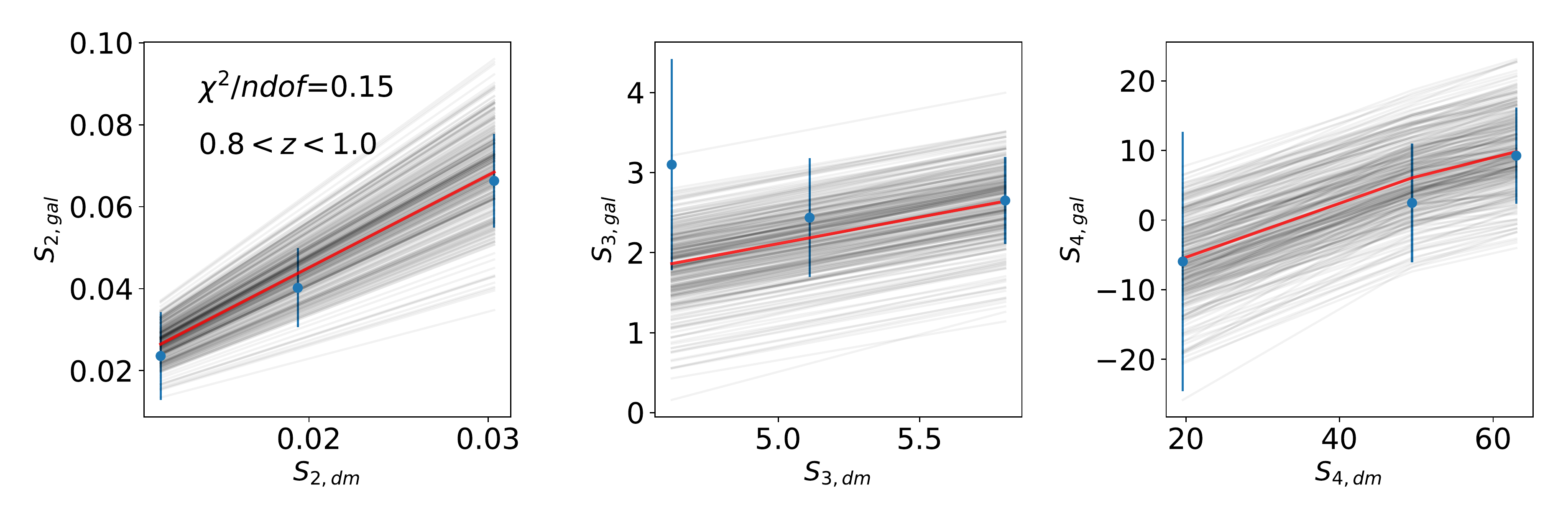}
\includegraphics[scale=0.35]{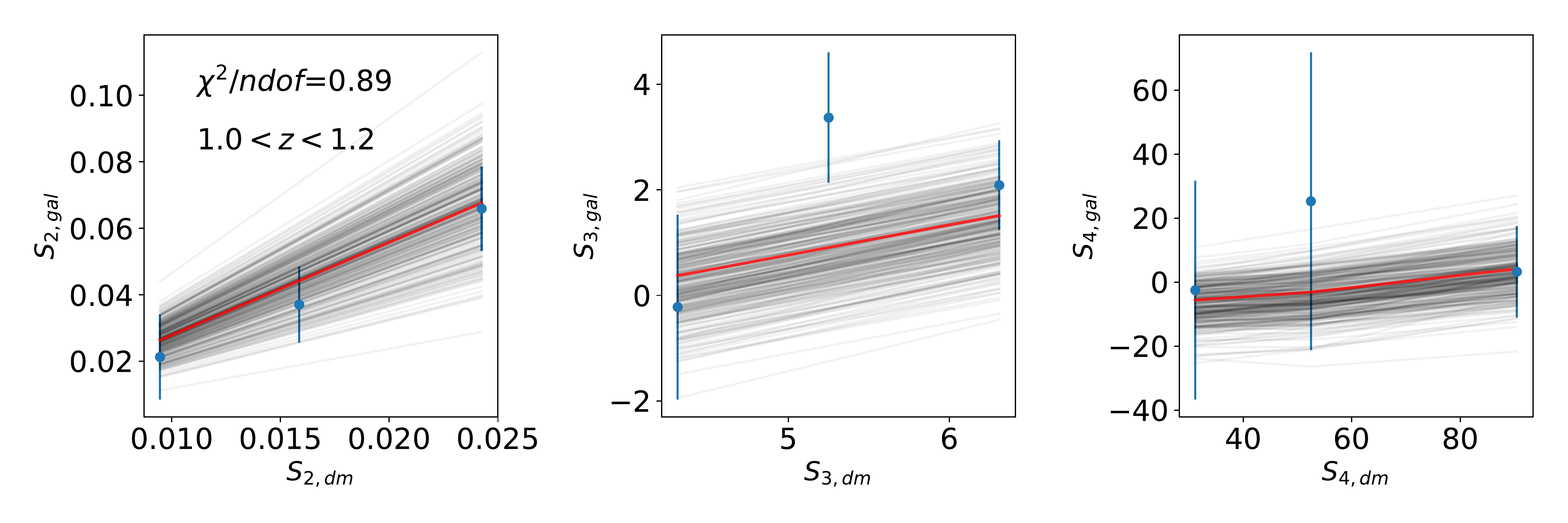}
\caption{Non-linear bias fits for DES-SV data. See caption in Figure \ref{fig:fits_mice_gaussian_nl} for more details.}
\label{fig:fits_data_gaussian_nl}
\end{figure*}
%\begin{figure*}
%\centering
%\includegraphics[scale=0.45,angle=90,trim={6cm 0.3cm 6cm 0.3cm},clip]{mcmc/fit_bin_3_tpz.pdf}
%\includegraphics[scale=0.45,angle=90,trim={6cm 0.3cm 6cm 0.3cm},clip]{mcmc/fit_bin_4_tpz.pdf}
%\caption{Non-linear bias fits for DES-SV data. See caption in Figure \ref{fig:fits_mice_gaussian_nl} for more details.}
%\label{fig:fits_data_gaussian_nl_2}
%\end{figure*}
\newpage
%-------------------------------------------------------------------------------------------------
%------------------------------------------------------------------------------------------------
\section*
{Affiliations}
$^{1}$ Instituto de Fisica Teorica UAM/CSIC, Universidad Autonoma de Madrid, 28049 Madrid, Spain\\
$^{2}$ Department of Physics and Astronomy, University of California, Irvine, 92602, USA\\
$^{3}$ Centro de Investigaciones Energ\'eticas, Medioambientales y Tecnol\'ogicas (CIEMAT), Madrid, Spain\\
$^{4}$ Department of Physics and Astronomy, University of California, Riverside, 92521, USA\\
$^{5}$ Kavli Institute for Cosmological Physics, University of Chicago, Chicago, IL 60637, USA\\
$^{6}$ DEDIP/DAP, IRFU, CEA, Universit\'e Paris-Saclay, F-91191 Gif-sur- Yvette, France\\
$^{7}$ Universit\'e Paris Diderot, AIM, Sorbonne Paris Cit\'e, CEA, CNRS, F-91191 Gif-sur-Yvette, France\\
$^{8}$ Fermi National Accelerator Laboratory, P. O. Box 500, Batavia, IL 60510, USA\\
$^{9}$ Institut d'Estudis Espacials de Catalunya (IEEC), 08193 Barcelona, Spain\\
$^{10}$ Institute of Space Sciences (ICE, CSIC),  Campus UAB, Carrer de Can Magrans, s/n,  08193 Barcelona, Spain\\
$^{11}$ Center for Cosmology and AstroParticle Physics, Department of Physics, The Ohio State University, 191 W Woodruff Ave,Columbus OH 43210, U.S.A.\\
$^{12}$ Cerro Tololo Inter-American Observatory, National Optical Astronomy Observatory, Casilla 603, La Serena, Chile\\
$^{13}$ Institute of Cosmology \& Gravitation, University of Portsmouth, Portsmouth, PO1 3FX, UK\\
$^{14}$ CNRS, UMR 7095, Institut d'Astrophysique de Paris, F-75014, Paris, France\\
$^{15}$ Sorbonne Universit\'es, UPMC Univ Paris 06, UMR 7095, Institut d'Astrophysique de Paris, F-75014, Paris, France\\
$^{16}$ Department of Physics \& Astronomy, University College London, Gower Street, London, WC1E 6BT, UK\\
$^{17}$ Kavli Institute for Particle Astrophysics \& Cosmology, P. O. Box 2450, Stanford University, Stanford, CA 94305, USA\\
$^{18}$ SLAC National Accelerator Laboratory, Menlo Park, CA 94025, USA\\
$^{19}$ Laborat\'orio Interinstitucional de e-Astronomia - LIneA, Rua Gal. Jos\'e Cristino 77, Rio de Janeiro, RJ - 20921-400, Brazil\\
$^{20}$ Observat\'orio Nacional, Rua Gal. Jos\'e Cristino 77, Rio de Janeiro, RJ - 20921-400, Brazil\\
$^{21}$ Department of Astronomy, University of Illinois at Urbana-Champaign, 1002 W. Green Street, Urbana, IL 61801, USA\\
$^{22}$ National Center for Supercomputing Applications, 1205 West Clark St., Urbana, IL 61801, USA\\
$^{23}$ Institut de F\'{\i}sica d'Altes Energies (IFAE), The Barcelona Institute of Science and Technology, Campus UAB, 08193 Bellaterra (Barcelona) Spain\\
$^{24}$ Department of Astronomy, University of Michigan, Ann Arbor, MI 48109, USA\\
$^{25}$ Department of Physics, University of Michigan, Ann Arbor, MI 48109, USA\\
$^{26}$ Department of Physics, ETH Zurich, Wolfgang-Pauli-Strasse 16, CH-8093 Zurich, Switzerland\\
$^{27}$ Santa Cruz Institute for Particle Physics, Santa Cruz, CA 95064, USA\\
$^{28}$ Harvard-Smithsonian Center for Astrophysics, Cambridge, MA 02138, USA\\
$^{29}$ Australian Astronomical Observatory, North Ryde, NSW 2113, Australia\\
$^{30}$ Departamento de F\'isica Matem\'atica, Instituto de F\'isica, Universidade de S\~ao Paulo, CP 66318, S\~ao Paulo, SP, 05314-970, Brazil\\
$^{31}$ Department of Physics and Astronomy, University of Pennsylvania, Philadelphia, PA 19104, USA\\
$^{32}$ George P. and Cynthia Woods Mitchell Institute for Fundamental Physics and Astronomy, and Department of Physics and Astronomy, Texas A\&M University, College Station, TX 77843,  USA\\
$^{33}$ Instituci\'o Catalana de Recerca i Estudis Avan\c{c}ats, E-08010 Barcelona, Spain\\
$^{34}$ Department of Physics and Astronomy, Pevensey Building, University of Sussex, Brighton, BN1 9QH, UK\\
$^{35}$ School of Physics and Astronomy, University of Southampton,  Southampton, SO17 1BJ, UK\\
$^{36}$ Brandeis University, Physics Department, 415 South Street, Waltham MA 02453\\
$^{37}$ Instituto de F\'isica Gleb Wataghin, Universidade Estadual de Campinas, 13083-859, Campinas, SP, Brazil\\
$^{38}$ Computer Science and Mathematics Division, Oak Ridge National Laboratory, Oak Ridge, TN 37831\\
$^{39}$ Argonne National Laboratory, 9700 South Cass Avenue, Lemont, IL 60439, USA\\

\section*{Acknowledgements}\label{sec:acknowledgements}  

%We are very grateful to our DES internal reviewers E. Gazta\~{n}aga, I. Sevilla, and A. J. Ross for their valuable comments that greatly improved the quality of this work. We thank Santiago \'Avila for useful discussions and providing catalogs for preliminary tests. 

We would like to thank Jonathan Loveday for carefully reading the manuscript, providing invaluable feedback that improved the overall quality of this work. We thank Anna M. Porredon for providing theoretical correlation functions for checking purposes and Cora Uhlemann for her insightful comments.

%We also thank T. H. Diehl, D. Gruen, and D. Burke for their useful comments. 

We acknowledge the use of data from the MICE simulations, publicly available at \url{http://www.ice.cat/mice}. We also acknowledge the use of Scipy, Numpy, Astropy, Healpy, emcee and iPython for this work.
FJS acknowledges support from the U.S. Department of Energy. Funding for the DES Projects has been provided by the U.S. Department of Energy, the U.S. National Science Foundation, the Ministry of Science and Education of Spain, 
the Science and Technology Facilities Council of the United Kingdom, the Higher Education Funding Council for England, the National Center for Supercomputing 
Applications at the University of Illinois at Urbana-Champaign, the Kavli Institute of Cosmological Physics at the University of Chicago, 
the Center for Cosmology and Astro-Particle Physics at the Ohio State University,
the Mitchell Institute for Fundamental Physics and Astronomy at Texas A\&M University, Financiadora de Estudos e Projetos, 
Funda{\c c}{\~a}o Carlos Chagas Filho de Amparo {\`a} Pesquisa do Estado do Rio de Janeiro, Conselho Nacional de Desenvolvimento Cient{\'i}fico e Tecnol{\'o}gico and 
the Minist{\'e}rio da Ci{\^e}ncia, Tecnologia e Inova{\c c}{\~a}o, the Deutsche Forschungsgemeinschaft and the Collaborating Institutions in the Dark Energy Survey. 

The Collaborating Institutions are Argonne National Laboratory, the University of California at Santa Cruz, the University of Cambridge, Centro de Investigaciones Energ{\'e}ticas, 
Medioambientales y Tecnol{\'o}gicas-Madrid, the University of Chicago, University College London, the DES-Brazil Consortium, the University of Edinburgh, 
the Eidgen{\"o}ssische Technische Hochschule (ETH) Z{\"u}rich, 
Fermi National Accelerator Laboratory, the University of Illinois at Urbana-Champaign, the Institut de Ci{\`e}ncies de l'Espai (IEEC/CSIC), 
the Institut de F{\'i}sica d'Altes Energies, Lawrence Berkeley National Laboratory, the Ludwig-Maximilians Universit{\"a}t M{\"u}nchen and the associated Excellence Cluster Universe, 
the University of Michigan, the National Optical Astronomy Observatory, the University of Nottingham, The Ohio State University, the University of Pennsylvania, the University of Portsmouth, 
SLAC National Accelerator Laboratory, Stanford University, the University of Sussex, Texas A\&M University, and the OzDES Membership Consortium.

Based in part on observations at Cerro Tololo Inter-American Observatory, National Optical Astronomy Observatory, which is operated by the Association of 
Universities for Research in Astronomy (AURA) under a cooperative agreement with the National Science Foundation.

The DES data management system is supported by the National Science Foundation under Grant Numbers AST-1138766 and AST-1536171.
The DES participants from Spanish institutions are partially supported by MINECO under grants AYA2015-71825, ESP2015-66861, FPA2015-68048, SEV-2016-0588, SEV-2016-0597, and MDM-2015-0509, 
some of which include ERDF funds from the European Union. IFAE is partially funded by the CERCA program of the Generalitat de Catalunya.
Research leading to these results has received funding from the European Research
Council under the European Union's Seventh Framework Program (FP7/2007-2013) including ERC grant agreements 240672, 291329, and 306478.
We  acknowledge support from the Australian Research Council Centre of Excellence for All-sky Astrophysics (CAASTRO), through project number CE110001020, and the Brazilian Instituto Nacional de Ci\^encia
e Tecnologia (INCT) e-Universe (CNPq grant 465376/2014-2).

This manuscript has been authored by Fermi Research Alliance, LLC under Contract No. DE-AC02-07CH11359 with the U.S. Department of Energy, Office of Science, Office of High Energy Physics. The United States Government retains and the publisher, by accepting the article for publication, acknowledges that the United States Government retains a non-exclusive, paid-up, irrevocable, world-wide license to publish or reproduce the published form of this manuscript, or allow others to do so, for United States Government purposes.

We are grateful for the extraordinary contributions of our CTIO colleagues and the DECam Construction, Commissioning and Science Verification
teams in achieving the excellent instrument and telescope conditions that have made this work possible.  The success of this project also 
relies critically on the expertise and dedication of the DES Data Management group.

%--------------------------------------------------------------------------------------
% the bibliography
\bibliography{des_cic_sv}

\begin{thebibliography}{}

\bibitem[\protect\citeauthoryear{Ade et~al.,}{Ade  et~al.}{2014}]{Ade:2013zuv}
Ade P. A.~R.,  et~al., 2014, Astron. Astrophys., 571, A16

\bibitem[\protect\citeauthoryear{{Amiaux}, {Scaramella} \& {Mellier}}{{Amiaux}
  et~al.}{2012}]{2012SPIE.8442E..0ZA}
{Amiaux} J.,  {Scaramella} R.,    {Mellier} 2012, in Space Telescopes and
  Instrumentation 2012: Optical, Infrared, and Millimeter Wave Vol.~8442 of
  Proc. SPIE, {Euclid mission: building of a reference survey}.
p. 84420Z

\bibitem[\protect\citeauthoryear{{Bardeen}, {Bond}, {Kaiser} \&
  {Szalay}}{{Bardeen} et~al.}{1986}]{1986ApJ...304...15B}
{Bardeen} J.,  {Bond} J.,  {Kaiser} N.,    {Szalay} A.,  1986, Astrophys. J.,
  304, 15

\bibitem[\protect\citeauthoryear{Bel, Hoffmann \& Gaztañaga}{Bel
  et~al.}{2015}]{Bel:2015jla}
Bel J.,  Hoffmann K.,    Gaztañaga E.,  2015, Mon. Not. Roy. Astron. Soc.,
  453, 259

\bibitem[\protect\citeauthoryear{Benitez}{Benitez}{2000}]{Benitez:1998br}
Benitez N.,  2000, Astrophys. J., 536, 571

\bibitem[\protect\citeauthoryear{Bernardeau}{Bernardeau}{1994}]{Bernardeau:1993qu}
Bernardeau F.,  1994, Astrophys. J., 433, 1

\bibitem[\protect\citeauthoryear{Bernardeau}{Bernardeau}{1996}]{Bernardeau:1995ty}
Bernardeau F.,  1996, Astron. Astrophys., 312, 11

\bibitem[\protect\citeauthoryear{{Bertin} \& {Arnouts}}{{Bertin} \&
  {Arnouts}}{1996}]{Bertin1996}
{Bertin} E.,  {Arnouts} S.,  1996, Astronomy and Astrophysics, Supplement, 117,
  393

\bibitem[\protect\citeauthoryear{{Blanton}, {Cen}, {Ostriker}, {Strauss} \&
  {Tegmark}}{{Blanton} et~al.}{2000}]{2000ApJ...531....1B}
{Blanton} M.,  {Cen} R.,  {Ostriker} J.~P.,  {Strauss} M.~A.,    {Tegmark} M.,
  2000, Astrophys. J., 531, 1

\bibitem[\protect\citeauthoryear{Bouchet, Juszkiewicz, Colombi \&
  Pellat}{Bouchet et~al.}{1992}]{Bouchet:1992uh}
Bouchet F.~R.,  Juszkiewicz R.,  Colombi S.,    Pellat R.,  1992, Astrophys.
  J., 394, L5

\bibitem[\protect\citeauthoryear{Carrasco~Kind \& Brunner}{Carrasco~Kind \&
  Brunner}{2013}]{Kind:2013eka}
Carrasco~Kind M.,  Brunner R.~J.,  2013, Mon. Not. Roy. Astron. Soc., 432, 1483

\bibitem[\protect\citeauthoryear{{Carrasco Kind} \& {Brunner}}{{Carrasco Kind}
  \& {Brunner}}{2013}]{2013MNRAS.432.1483C}
{Carrasco Kind} M.,  {Brunner} R.~J.,  2013, MNRAS, 432, 1483

\bibitem[\protect\citeauthoryear{Chang, Pujol, {Gazta{\~n}aga}, {Amara},
  R{\'e}fr{\'e}gier, Bacon et~al.,}{Chang et~al.}{2016}]{2016MNRAS.459.3203C}
Chang C.,  Pujol A.,  {Gazta{\~n}aga} E.,  {Amara} A.,  R{\'e}fr{\'e}gier A.,
  Bacon D.,    et~al., 2016, MNRAS, 459, 3203

\bibitem[\protect\citeauthoryear{Clerkin et~al.,}{Clerkin
  et~al.}{2016}]{Clerkin:2016kyr}
Clerkin L.,  et~al., 2016, Mon. Not. Roy. Astron. Soc.

\bibitem[\protect\citeauthoryear{{Crocce}, {Carretero}, {Bauer}, {Ross},
  {Sevilla-Noarbe}, {Giannantonio}, {Sobreira}, {Sanchez}, {Gazta{\~n}aga} \&
  {DES Collaboration}}{{Crocce} et~al.}{2016}]{2016MNRAS.455.4301C}
{Crocce} M.,  {Carretero} J.,  {Bauer} A.~H.,  {Ross} A.~J.,  {Sevilla-Noarbe}
  I.,  {Giannantonio} T.,  {Sobreira} F.,  {Sanchez} J.,  {Gazta{\~n}aga} E.,
   {DES Collaboration} 2016, MNRAS, 455, 4301

\bibitem[\protect\citeauthoryear{Crocce, Castander, Gazta{\~n}aga, Fosalba \&
  Carretero}{Crocce et~al.}{2015}]{Crocce:2013vda}
Crocce M.,  Castander F.~J.,  Gazta{\~n}aga E.,  Fosalba P.,    Carretero J.,
  2015, Mon. Not. Roy. Astron. Soc., 453, 1513

\bibitem[\protect\citeauthoryear{{Crocce}, {Fosalba}, {Castander} \&
  {Gazta{\~n}aga}}{{Crocce} et~al.}{2010}]{2010MNRAS.403.1353C}
{Crocce} M.,  {Fosalba} P.,  {Castander} F.~J.,    {Gazta{\~n}aga} E.,  2010,
  MNRAS, 403, 1353

\bibitem[\protect\citeauthoryear{{Dark Energy Survey Collaboration}
  et~al.,}{{Dark Energy Survey Collaboration}
  et~al.}{2016}]{2016MNRAS.460.1270D}
{Dark Energy Survey Collaboration} et~al., 2016, MNRAS, 460, 1270

\bibitem[\protect\citeauthoryear{{Davis}, {Geller} \& {Huchra}}{{Davis}
  et~al.}{1978}]{1978ApJ...221....1D}
{Davis} M.,  {Geller} M.~J.,    {Huchra} J.,  1978, Astrophys. J., 221, 1

\bibitem[\protect\citeauthoryear{{Dressler}}{{Dressler}}{1980}]{1980ApJ...236..351D}
{Dressler} A.,  1980, Astrophys. J., 236, 351

\bibitem[\protect\citeauthoryear{{Drlica-Wagner} et~al.,}{{Drlica-Wagner}
  et~al.}{2018}]{2018ApJS..235...33D}
{Drlica-Wagner} A.,  et~al., 2018, ApJS, 235, 33

\bibitem[\protect\citeauthoryear{Efron}{Efron}{1979}]{Efron79}
Efron B.,  1979, The Annals of Statistics, 7, 1

\bibitem[\protect\citeauthoryear{{Efstathiou}, {Kaiser}, {Saunders},
  {Lawrence}, {Rowan-Robinson}, {Ellis} \& {Frenk}}{{Efstathiou}
  et~al.}{1990}]{1990MNRAS.247P..10E}
{Efstathiou} G.,  {Kaiser} N.,  {Saunders} W.,  {Lawrence} A.,
  {Rowan-Robinson} M.,  {Ellis} R.~S.,    {Frenk} C.~S.,  1990, MNRAS, 247, 10P

\bibitem[\protect\citeauthoryear{Eriksen \& Gaztanaga}{Eriksen \&
  Gaztanaga}{2015}]{Eriksen:2015hqa}
Eriksen M.,  Gaztanaga E.,  2015

\bibitem[\protect\citeauthoryear{{Foreman-Mackey}, {Hogg}, {Lang} \&
  {Goodman}}{{Foreman-Mackey} et~al.}{2013}]{2013PASP..125..306F}
{Foreman-Mackey} D.,  {Hogg} D.~W.,  {Lang} D.,    {Goodman} J.,  2013, PASP,
  125, 306

\bibitem[\protect\citeauthoryear{{Fosalba}, {Gazta{\~n}aga}, {Castander} \&
  {Manera}}{{Fosalba} et~al.}{2008}]{2008MNRAS.391..435F}
{Fosalba} P.,  {Gazta{\~n}aga} E.,  {Castander} F.~J.,    {Manera} M.,  2008,
  MNRAS, 391, 435

\bibitem[\protect\citeauthoryear{{Frieman} \& {Gazta{\~n}aga}}{{Frieman} \&
  {Gazta{\~n}aga}}{1999}]{friemannGaz}
{Frieman} J.~A.,  {Gazta{\~n}aga} E.,  1999, Astrophys. J., 521, L83

\bibitem[\protect\citeauthoryear{Fry}{Fry}{1996}]{Fryevolution}
Fry J.~N.,  1996, The Astrophysical Journal Letters, 461, L65

\bibitem[\protect\citeauthoryear{Fry \& Gazta{\~n}aga}{Fry \&
  Gazta{\~n}aga}{1993}]{Fry:1992vr}
Fry J.~N.,  Gazta{\~n}aga E.,  1993, Astrophys. J., 413, 447

\bibitem[\protect\citeauthoryear{{Garcia-Fernandez}, {Sanchez},
  {Sevilla-Noarbe}, {Suchyta}, {Huff}, {Gaztanaga}, {Aleksi{\'c}}, {Ponce},
  {Castander}, {Hoyle}, {Abbott}, {Abdalla} et~al.,}{{Garcia-Fernandez}
  et~al.}{2018}]{2018MNRAS.476.1071G}
{Garcia-Fernandez} M.,  {Sanchez} E.,  {Sevilla-Noarbe} I.,  {Suchyta} E.,
  {Huff} E.~M.,  {Gaztanaga} E.,  {Aleksi{\'c}} J.,  {Ponce} R.,  {Castander}
  F.~J.,  {Hoyle} B.,  {Abbott} T.~M.~C.,  {Abdalla} F.~B.,    et~al., 2018,
  MNRAS, 476, 1071

\bibitem[\protect\citeauthoryear{Gazta{\~n}aga}{Gazta{\~n}aga}{1994}]{Gaztanaga:1993ru}
Gazta{\~n}aga E.,  1994, Mon. Not. Roy. Astron. Soc., 268, 913

\bibitem[\protect\citeauthoryear{Gaztanaga, Eriksen, Crocce, Castander, Fosalba
  et~al.,}{Gaztanaga et~al.}{2011}]{Gaztanaga:2011yi}
Gaztanaga E.,  Eriksen M.,  Crocce M.,  Castander F.,  Fosalba P.,    et~al.,
  2011

\bibitem[\protect\citeauthoryear{{Giannantonio}, {Fosalba}, {Cawthon}, {Omori}
  \& et al.}{{Giannantonio} et~al.}{2016}]{2016MNRAS.456.3213G}
{Giannantonio} T.,  {Fosalba} P.,  {Cawthon} R.,  {Omori} Y.,    et al. 2016,
  MNRAS, 456, 3213

\bibitem[\protect\citeauthoryear{Gil-Mar\'{i}n, Noreña, Verde, Percival,
  Wagner, Manera \& Schneider}{Gil-Mar\'{i}n et~al.}{2015}]{Gil-Marin:2014sta}
Gil-Mar\'{i}n H.,  Noreña J.,  Verde L.,  Percival W.~J.,  Wagner C.,  Manera
  M.,    Schneider D.~P.,  2015, Mon. Not. Roy. Astron. Soc., 451, 539

\bibitem[\protect\citeauthoryear{{G{\'o}rski}, {Hivon}, {Banday}, {Wandelt},
  {Hansen}, {Reinecke} \& {Bartelmann}}{{G{\'o}rski} et~al.}{2005}]{Gorski2005}
{G{\'o}rski} K.~M.,  {Hivon} E.,  {Banday} A.~J.,  {Wandelt} B.~D.,  {Hansen}
  F.~K.,  {Reinecke} M.,    {Bartelmann} M.,  2005, ApJ, 622, 759

\bibitem[\protect\citeauthoryear{{Gruen}, {Friedrich}, {Krause}, {DeRose},
  {Cawthon}, {Davis}, {Elvin-Poole}, {Rykoff} et~al.,}{{Gruen}
  et~al.}{2017}]{2017arXiv171005045G}
{Gruen} D.,  {Friedrich} O.,  {Krause} E.,  {DeRose} J.,  {Cawthon} R.,
  {Davis} C.,  {Elvin-Poole} J.,  {Rykoff} E.~S.,    et~al., 2017, ArXiv
  e-prints

\bibitem[\protect\citeauthoryear{{Heath}}{{Heath}}{1977}]{Heath1977}
{Heath} D.~J.,  1977, MNRAS, 179, 351

\bibitem[\protect\citeauthoryear{Hoffmann, Bel \& Gaztanaga}{Hoffmann
  et~al.}{2015}]{Hoffmann:2015mma}
Hoffmann K.,  Bel J.,    Gaztanaga E.,  2015, Mon. Not. Roy. Astron. Soc., 450,
  1674

\bibitem[\protect\citeauthoryear{{Ivezi{\'c}}, {Connolly}, {Vanderplas} \&
  {Gray}}{{Ivezi{\'c}} et~al.}{2014}]{astroMLText}
{Ivezi{\'c}} {\v Z}.,  {Connolly} A.,  {Vanderplas} J.,    {Gray} A.,  2014,
  Statistics, Data Mining and Machine Learning in Astronomy.
Princeton University Press

\bibitem[\protect\citeauthoryear{{Ivezi{\'c}}, {Kahn}, {Tyson}, {Abel},
  {Acosta}, {Allsman}, {Alonso}, {AlSayyad}, {Anderson}, {Andrew} \& et
  al.}{{Ivezi{\'c}} et~al.}{2008}]{2008arXiv0805.2366I}
{Ivezi{\'c}} {\v Z}.,  {Kahn} S.~M.,  {Tyson} J.~A.,  {Abel} B.,  {Acosta} E.,
  {Allsman} R.,  {Alonso} D.,  {AlSayyad} Y.,  {Anderson} S.~F.,  {Andrew} J.,
    et al. 2008, ArXiv e-prints

\bibitem[\protect\citeauthoryear{Kaiser}{Kaiser}{1984}]{Kaiser:1984sw}
Kaiser N.,  1984, Astrophys. J., 284, L9

\bibitem[\protect\citeauthoryear{{Kollmeier}, {Zasowski} \& {Rix}}{{Kollmeier}
  et~al.}{2017}]{2017arXiv171103234K}
{Kollmeier} J.~A.,  {Zasowski} G.,    {Rix} 2017, ArXiv e-prints

\bibitem[\protect\citeauthoryear{{Le F{\`e}vre}, {Cassata}, {Cucciati},
  {Garilli}, {Ilbert}, {Le Brun}, {Maccagni}, {Moreau}, {Scodeggio}
  et~al.,}{{Le F{\`e}vre} et~al.}{2013}]{2013A&A...559A..14L}
{Le F{\`e}vre} O.,  {Cassata} P.,  {Cucciati} O.,  {Garilli} B.,  {Ilbert} O.,
  {Le Brun} V.,  {Maccagni} D.,  {Moreau} C.,  {Scodeggio} M.,    et~al., 2013,
  Astronomy and Astrophysics, 559, A14

\bibitem[\protect\citeauthoryear{{Leicht}, {Uhlemann}, {Villaescusa-Navarro},
  {Codis}, {Hernquist} \& {Genel}}{{Leicht} et~al.}{2018}]{2018arXiv180809968L}
{Leicht} O.,  {Uhlemann} C.,  {Villaescusa-Navarro} F.,  {Codis} S.,
  {Hernquist} L.,    {Genel} S.,  2018, ArXiv e-prints

\bibitem[\protect\citeauthoryear{{Leistedt}, {Peiris}, {Elsner}, {Benoit-Levy},
  {Amara} et~al.,}{{Leistedt} et~al.}{2016}]{2016ApJS..226...24L}
{Leistedt} B.,  {Peiris} H.,  {Elsner} F.,  {Benoit-Levy} A.,  {Amara} A.,
  et~al., 2016, The Astrophysical Journal Suplement, 226, 24

\bibitem[\protect\citeauthoryear{{Lilly}, {Le Brun}, {Maier}, {Mainieri},
  {Mignoli}, {Scodeggio}, {Zamorani}, {Carollo}, {Contini}, {Kneib}, {Le
  F{\`e}vre}, {Renzini}, {Bardelli}, {Bolzonella}, {Bongiorno}, {Caputi}
  et~al.,}{{Lilly} et~al.}{2009}]{2009ApJS..184..218L}
{Lilly} S.~J.,  {Le Brun} V.,  {Maier} C.,  {Mainieri} V.,  {Mignoli} M.,
  {Scodeggio} M.,  {Zamorani} G.,  {Carollo} M.,  {Contini} T.,  {Kneib} J.-P.,
   {Le F{\`e}vre} O.,  {Renzini} A.,  {Bardelli} S.,  {Bolzonella} M.,
  {Bongiorno} A.,  {Caputi} K.,    et~al., 2009, ApJS, 184, 218

\bibitem[\protect\citeauthoryear{{Lilly}, {Le F{\`e}vre}, {Renzini},
  {Zamorani}, {Scodeggio}, {Contini}, {Carollo}, {Hasinger}, {Kneib}, {Iovino},
  {Le Brun}, {Maier}, {Mainieri}, {Mignoli}, {Silverman} et~al.,}{{Lilly}
  et~al.}{2007}]{2007ApJS..172...70L}
{Lilly} S.~J.,  {Le F{\`e}vre} O.,  {Renzini} A.,  {Zamorani} G.,  {Scodeggio}
  M.,  {Contini} T.,  {Carollo} C.~M.,  {Hasinger} G.,  {Kneib} J.-P.,
  {Iovino} A.,  {Le Brun} V.,  {Maier} C.,  {Mainieri} V.,  {Mignoli} M.,
  {Silverman} J.,    et~al., 2007, ApJS, 172, 70

\bibitem[\protect\citeauthoryear{{Manera} \& {Gazta{\~n}aga}}{{Manera} \&
  {Gazta{\~n}aga}}{2011}]{2011MNRAS.415..383M}
{Manera} M.,  {Gazta{\~n}aga} E.,  2011, Mon. Not. Roy. Astron. Soc., 415, 383

\bibitem[\protect\citeauthoryear{Manera, Sheth \& Scoccimarro}{Manera
  et~al.}{2010}]{Manera:2009ak}
Manera M.,  Sheth R.~K.,    Scoccimarro R.,  2010, Mon. Not. Roy. Astron. Soc.,
  402, 589

\bibitem[\protect\citeauthoryear{{Masci} \& {SWIRE Team}}{{Masci} \& {SWIRE
  Team}}{2006}]{2006ASPC..357..271M}
{Masci} F.~J.,  {SWIRE Team} 2006, in {Armus} L.,  {Reach} W.~T.,  eds,
  Astronomical Society of the Pacific Conference Series Vol.~357 of
  Astronomical Society of the Pacific Conference Series, {Large Scale Structure
  at 24 Microns in the SWIRE Survey}.
p.~271

\bibitem[\protect\citeauthoryear{Mo \& White}{Mo \& White}{1996}]{Mo:1995cs}
Mo H.~J.,  White S. D.~M.,  1996, Mon. Not. Roy. Astron. Soc., 282, 347

\bibitem[\protect\citeauthoryear{{Monet}, {Levine}, {Canzian}, {Ables}, {Bird},
  {Dahn}, {Guetter}, {Harris} et~al.,}{{Monet}
  et~al.}{2003}]{2003AJ....125..984M}
{Monet} D.,  {Levine} S.,  {Canzian} B.,  {Ables} H.,  {Bird} A.,  {Dahn} C.,
  {Guetter} H.,  {Harris} H.,    et~al., 2003, The Astrophysical Journal, 125,
  984

\bibitem[\protect\citeauthoryear{{Norberg}, {Baugh}, {Hawkins} \&
  {Maddox}}{{Norberg} et~al.}{2002}]{2002MNRAS.332..827N}
{Norberg} P.,  {Baugh} C.,  {Hawkins} E.,    {Maddox} S.,  2002, Mon. Not. Roy.
  Astron. Soc., 332, 827

\bibitem[\protect\citeauthoryear{{Norberg}, {Baugh}, {Gazta{\~n}aga} \&
  {Croton}}{{Norberg} et~al.}{2009}]{2009MNRAS.396...19N}
{Norberg} P.,  {Baugh} C.~M.,  {Gazta{\~n}aga} E.,    {Croton} D.~J.,  2009,
  MNRAS, 396, 19

\bibitem[\protect\citeauthoryear{Nusser \& Davis}{Nusser \&
  Davis}{1994}]{Nusser:1993sx}
Nusser A.,  Davis M.,  1994, Astrophys. J., 421, L1

\bibitem[\protect\citeauthoryear{{Peebles}}{{Peebles}}{1980}]{Peebles1980}
{Peebles} P.~J.~E.,  1980, {The large-scale structure of the universe}.
Princeton University Press

\bibitem[\protect\citeauthoryear{{Pen}}{{Pen}}{1998}]{1998ApJ...504..601P}
{Pen} U.-L.,  1998, The Astrophysical Journal, 504, 601

\bibitem[\protect\citeauthoryear{Pollack, Smith \& Porciani}{Pollack
  et~al.}{2014}]{Pollack:2013alj}
Pollack J.~E.,  Smith R.~E.,    Porciani C.,  2014, Mon. Not. Roy. Astron.
  Soc., 440, 555

\bibitem[\protect\citeauthoryear{{Prat}, {S{\'a}nchez}, {Miquel}, {Kwan},
  {Blazek}, {Bonnett}, {Amara}, {Bridle}, {Clampitt}, {Crocce}, {Fosalba},
  {Gaztanaga} et~al.,}{{Prat} et~al.}{2018}]{2018MNRAS.473.1667P}
{Prat} J.,  {S{\'a}nchez} C.,  {Miquel} R.,  {Kwan} J.,  {Blazek} J.,
  {Bonnett} C.,  {Amara} A.,  {Bridle} S.~L.,  {Clampitt} J.,  {Crocce} M.,
  {Fosalba} P.,  {Gaztanaga} E.,    et~al., 2018, MNRAS, 473, 1667

\bibitem[\protect\citeauthoryear{{Press} \& {Schechter}}{{Press} \&
  {Schechter}}{1974}]{PressSchechter}
{Press} W.~H.,  {Schechter} P.,  1974, Astrophys. J., 187, 425

\bibitem[\protect\citeauthoryear{Pujol, Hoffmann, Jiménez \& Gaztañaga}{Pujol
  et~al.}{2017}]{Pujol:2015wna}
Pujol A.,  Hoffmann K.,  Jiménez N.,    Gaztañaga E.,  2017, Astron.
  Astrophys., 598, A103

\bibitem[\protect\citeauthoryear{Ross, Brunner \& Myers}{Ross
  et~al.}{2006}]{Ross:2006zr}
Ross A.~J.,  Brunner R.~J.,    Myers A.~D.,  2006, Astrophys. J., 649, 48

\bibitem[\protect\citeauthoryear{{Rozo}, {Rykoff}, {Abate}, {Bonnett},
  {Crocce}, {Davis}, {Hoyle}, {Leistedt}, {Peiris}, {Wechsler}, {Abbott},
  {Abdalla}, {Banerji} et~al.,}{{Rozo} et~al.}{2016}]{2016MNRAS.461.1431R}
{Rozo} E.,  {Rykoff} E.~S.,  {Abate} A.,  {Bonnett} C.,  {Crocce} M.,  {Davis}
  C.,  {Hoyle} B.,  {Leistedt} B.,  {Peiris} H.~V.,  {Wechsler} R.~H.,
  {Abbott} T.,  {Abdalla} F.~B.,  {Banerji} M.,    et~al., 2016, MNRAS, 461,
  1431

\bibitem[\protect\citeauthoryear{{S{\'a}nchez}, {Carrasco Kind}, {Lin},
  {Miquel}, {Abdalla}, {Amara}, {Banerji}, {Bonnett}, {Brunner}, {Capozzi},
  {Carnero}, {Castander}, {da Costa}, {Cunha} et~al.,}{{S{\'a}nchez}
  et~al.}{2014}]{2014MNRAS.445.1482S}
{S{\'a}nchez} C.,  {Carrasco Kind} M.,  {Lin} H.,  {Miquel} R.,  {Abdalla}
  F.~B.,  {Amara} A.,  {Banerji} M.,  {Bonnett} C.,  {Brunner} R.,  {Capozzi}
  D.,  {Carnero} A.,  {Castander} F.~J.,  {da Costa} L.~A.~N.,  {Cunha} C.,
  et~al., 2014, MNRAS, 445, 1482

\bibitem[\protect\citeauthoryear{{S{\'a}nchez}, {Carnero},
  {Garc{\'{\i}}a-Bellido}, {Gazta{\~n}aga}, {de Simoni}, {Crocce}, {Cabr{\'e}},
  {Fosalba} \& {Alonso}}{{S{\'a}nchez} et~al.}{2011}]{2011MNRAS.411..277S}
{S{\'a}nchez} E.,  {Carnero} A.,  {Garc{\'{\i}}a-Bellido} J.,  {Gazta{\~n}aga}
  E.,  {de Simoni} F.,  {Crocce} M.,  {Cabr{\'e}} A.,  {Fosalba} P.,
  {Alonso} D.,  2011, MNRAS, 411, 277

\bibitem[\protect\citeauthoryear{{Sato} \& {Matsubara}}{{Sato} \&
  {Matsubara}}{2013}]{2013PhRvD..87l3523S}
{Sato} M.,  {Matsubara} T.,  2013, Physical Review Letters D, 87, 123523

\bibitem[\protect\citeauthoryear{Scherrer \& Weinberg}{Scherrer \&
  Weinberg}{1998}]{Scherrer:1997hp}
Scherrer R.~J.,  Weinberg D.~H.,  1998, Astrophys. J., 504, 607

\bibitem[\protect\citeauthoryear{{Schlegel}, {Finkbeiner} \&
  {Davis}}{{Schlegel} et~al.}{1998}]{1998ApJ...500..525S}
{Schlegel} D.,  {Finkbeiner} D.,    {Davis} M.,  1998, The Astrophysical
  Journal, 500, 525

\bibitem[\protect\citeauthoryear{Sheth \& Tormen}{Sheth \&
  Tormen}{1999}]{Sheth:1999mn}
Sheth R.~K.,  Tormen G.,  1999, Mon. Not. Roy. Astron. Soc., 308, 119

\bibitem[\protect\citeauthoryear{{Szapudi}}{{Szapudi}}{1998}]{1998ApJ...497...16S}
{Szapudi} I.,  1998, ApJ, 497, 16

\bibitem[\protect\citeauthoryear{{Szapudi}, {Frieman}, {Scoccimarro}, {Szalay},
  {Connolly}, {Dodelson}, {Eisenstein}, {Gunn}, {Johnston}, {Kent}, {Loveday},
  {Meiksin}, {Nichol} et~al.,}{{Szapudi} et~al.}{2002}]{2002ApJ...570...75S}
{Szapudi} I.,  {Frieman} J.~A.,  {Scoccimarro} R.,  {Szalay} A.~S.,  {Connolly}
  A.~J.,  {Dodelson} S.,  {Eisenstein} D.~J.,  {Gunn} J.~E.,  {Johnston} D.,
  {Kent} S.,  {Loveday} J.,  {Meiksin} A.,  {Nichol} R.~C.,    et~al., 2002,
  ApJ, 570, 75

\bibitem[\protect\citeauthoryear{Tegmark \& Peebles}{Tegmark \&
  Peebles}{1998}]{Tegmark:1998wm}
Tegmark M.,  Peebles P. J.~E.,  1998, Astrophys. J., 500, L79

\bibitem[\protect\citeauthoryear{Willmer, Maia, Mendes, Alonso, Rios, Chaves \&
  de Mello}{Willmer et~al.}{1999}]{Willmer:1999hf}
Willmer C. N.~A.,  Maia M. A.~G.,  Mendes S.~O.,  Alonso M.~V.,  Rios L.~A.,
  Chaves O.~L.,    de Mello D.~F.,  1999, Astron. J., 118, 1131

\bibitem[\protect\citeauthoryear{{Wolk}, {McCracken}, {Colombi}, {Fry},
  {Kilbinger}, {Hudelot}, {Mellier} \& {Ilbert}}{{Wolk}
  et~al.}{2013}]{2013MNRAS.435....2W}
{Wolk} M.,  {McCracken} H.~J.,  {Colombi} S.,  {Fry} J.~N.,  {Kilbinger} M.,
  {Hudelot} P.,  {Mellier} Y.,    {Ilbert} O.,  2013, MNRAS, 435, 2

\bibitem[\protect\citeauthoryear{{Yang} \& {Saslaw}}{{Yang} \&
  {Saslaw}}{2011}]{2011ApJ...729..123Y}
{Yang} A.,  {Saslaw} W.~C.,  2011, ApJ, 729, 123

\bibitem[\protect\citeauthoryear{{Zehavi}, {Blanton}, {Frieman} \&
  {Weinberg}}{{Zehavi} et~al.}{2002}]{2002ApJ...571..172Z}
{Zehavi} I.,  {Blanton} M.,  {Frieman} J.,    {Weinberg} D.,  2002, Astrophys.
  J., 571, 172

\end{thebibliography}
%\input des_cic_sv_biblio
%--------------------------------------------------------------------------------------
%\label{lastpage}
%%%%%%%%%%%%%%%%%%%%%%%%%%%%%%%%%%%%%%%%%%%%%%%%%%%%%%%%%%%%%%%%%%%%%%%%%%%%%%%%%%%%%%%
\end{document}